\documentclass[aps,prd,twocolumn,superscriptaddress,nofootinbib,10pt]{revtex4-2}
\usepackage{threeparttable}
\usepackage{amsmath,amssymb}
\usepackage{graphicx}
\usepackage{xcolor}
\usepackage{booktabs}
\usepackage{pifont}
\usepackage{tikz}
\usetikzlibrary{positioning,shapes.geometric,arrows.meta}
\usepackage{hyperref}
\hypersetup{
  pdftitle={Dissecting the Hubble tension: Insights from a diverse set of Sound-Horizon-Free \texorpdfstring{$H_0$}{H0} measurements},
  pdfauthor={Ioannis Pantos, Leandros Perivolaropoulos},
  colorlinks=true,  
  linkcolor=green!40!black,
  citecolor=blue,
  urlcolor=blue
}

\begin{document}

\title{Dissecting the Hubble tension: Insights from a diverse set of Sound-Horizon-Free \texorpdfstring{$H_0$}{H0} measurements}

\author{Ioannis Pantos}
\email{i.pantos@uoi.gr} % Optional: add email if desired
\affiliation{Department of Physics, University of Ioannina, Greece}

\author{Leandros Perivolaropoulos}
\email{leandros@uoi.gr} % Optional: add email if desired
\affiliation{Department of Physics, University of Ioannina, Greece}

\date{\today}

\begin{abstract}
The Hubble tension is commonly framed as a discrepancy between local, late-time measurements favoring $H_0 \approx 73$ km s$^{-1}$ Mpc$^{-1}$ and early-time, Sound-Horizon-based measurements favoring $H_0 \approx 67$ km s$^{-1}$ Mpc$^{-1}$. We challenge this viewpoint by analyzing 88 Sound Horizon Free $H_0$ measurements, categorized into four classes: Distance Ladder measurements using local calibrators; Local $\Lambda$CDM measurements assuming the standard expansion history; Pure Local measurements independent of $H(z)$ shape; and CMB Sound--Horizon--Free measurements using CMB data without the Sound Horizon scale.
Our analysis reveals that the 30 Distance Ladder measurements yield $H_0 = 72.73 \pm 0.39$ km s$^{-1}$ Mpc$^{-1}$ ($\chi^2_\nu = 0.72$), while the 58  Distance Ladder-Independent/Sound Horizon Free measurements collectively yield $H_0 = 69.37 \pm 0.34$ km s$^{-1}$ Mpc$^{-1}$ ($\chi^2_\nu = 0.95$), a $6.5\sigma$ tension exceeding the Planck--SH0ES discrepancy. This tension remains significant at a minimum value of $3.9\sigma$ after accounting for correlations. Among categories, Local $\Lambda$CDM measurements favor the lowest value ($H_0 = 67.61\pm 0.96$ km s$^{-1}$ Mpc$^{-1}$), Pure Local yield an intermediate value ($H_0 = 71.03 \pm 0.69$ km s$^{-1}$ Mpc$^{-1}$), and CMB Sound Horizon Free measurements give $H_0 = 69.07 \pm 0.44$ km s$^{-1}$ Mpc$^{-1}$.
We conclude that the Hubble tension is better characterized as a discrepancy between the Distance Ladder and all other methodologies, rather than an early-vs-late-time split. We also identify a $2.9\sigma$ internal tension among Distance Ladder Independent/Sound Horizon Free measurements (noting its sensitivity to categorization choices): analyses assuming $\Lambda$CDM systematically recover lower $H_0$ values by $\sim$3.4 km s$^{-1}$ Mpc$^{-1}$ compared to cosmological-model-independent methods. This suggests either unrecognized systematics in the Distance Ladder or deviations from $\Lambda$CDM or both.
\end{abstract}
\maketitle
%======================================================================
%\section{Introduction}
%\label{sec:introduction}
%======================================================================
% Goals for this section:
% 1. State the standard view: Tension = Local (High) vs Early/Sound Horizon (Low).
% 2. Introduce the counter-argument: Many local/late-time probes do not agree with Distance Ladder.
% 3. Define the scope: 88 measurements, strictly independent of r_s (sound horizon).
% 4. Outline the classification scheme (The 4 categories) which is the core of this paper.

%======================================================================
\section{Introduction}
\label{sec:introduction}
%======================================================================

The determination of the Hubble constant $H_0$ has reached a level of precision that has arguably transformed a measurement discrepancy into a crisis for the standard cosmological model. The ``Hubble tension'' refers to the persistent $>5\sigma$ disagreement between the value inferred from early-Universe probes and that measured from the late-Universe local Distance Ladder.

The standard view frames this tension as a dichotomy between ``early'' and ``late'' times. The Planck CMB analysis, assuming $\Lambda$CDM, yields $H_0 = 67.4 \pm 0.5\,\mathrm{km\,s^{-1}\,Mpc^{-1}}$~\cite{Planck2018}, an inference dependent on the Sound Horizon scale at the drag epoch ($r_s$). The local Distance Ladder, anchored by geometric measurements and extended via Cepheids to SNe~Ia, favors $H_0 = 73.04 \pm 1.04\,\mathrm{km\,s^{-1}\,Mpc^{-1}}$~\cite{Riess2022}, reinforced by JWST observations~\cite{Riess2025} and extended network analyses~\cite{Casertano2025}. If both are accurate, $\Lambda$CDM is incomplete, potentially necessitating new physics~\cite{DiValentino2021review,Knox2020,Abdalla:2022yfr,Perivolaropoulos:2021jda,Alestas2020,Turner:2025qmf,Chaudhary:2025uzr,Perivolaropoulos:2025zjt}. 

Various theoretical approaches have been proposed to address the Hubble tension, which can be broadly classified into three categories based on the cosmic epoch at which they introduce new physics \cite{DiValentino2021review,Knox2020}.

\textbf{Early-time models} attempt to resolve the tension by modifying conditions during the recombination epoch, thereby reducing the Sound Horizon scale $r_s$ that serves as the standard ruler for CMB-based $H_0$ inference. Prominent examples include Early Dark Energy (EDE) scenarios and modified gravity theories. However, these approaches face significant challenges: they typically fail to fully resolve the tension, often exacerbate the $S_8$ tension by enhancing matter clustering, and require considerable fine-tuning of model parameters~\cite{Ruchika:2025sbb,Smith:2025icl,Poulin2019,Kamionkowski2023,Smith2021,Jedamzik2021,Vagnozzi2020,Giare:2025ath}.

\textbf{Late-time models} propose deformations in the Hubble expansion history $H(z)$ at low redshifts $z \lesssim 2$, often through dynamical dark energy with an evolving equation of state~\cite{Plaza:2025nip,Cheng:2025lod,SanchezLopez:2025uzw,Park:2025fbl,Shajib:2025tpd,Escamilla:2024ahl,DiValentino2020a,DiValentino2020b,Lee:2025pzo,Wolf:2025jlc,Ferri:2025tuo}. To understand the implications, it is essential to first outline the theoretical classification of Dark Energy dynamics. The evolution of dark energy is characterized by its equation of state parameter, $w$, defined as the ratio of its pressure to its energy density ($w = P/\rho$).

In the standard $\Lambda$CDM model, this parameter corresponds to a Cosmological Constant and is strictly fixed at $w = -1$. However, dynamical dark energy models are generally categorized into two distinct regimes based on deviations from this value: Quintessence  ($-1 < w < -1/3$), which is typically modeled by a scalar field rolling down a potential, and the Phantom regime ($w < -1$), where the fluid exhibits super-negative pressure, leading to an increasing energy density over time~\cite{Ozulker:2025ehg,Silva:2025hxw}

A dynamical evolution in which the parameter $w(z)$ traverses the $w = -1$ boundary is referred to as Phantom Crossing. This transition presents significant theoretical challenges; within the framework of simple perfect fluids or single scalar field models, crossing this boundary typically triggers gradient instabilities or introduces ghost degrees of freedom. Consequently, viable phantom crossing scenarios usually necessitate frameworks beyond standard General Relativity, such as Modified Gravity theories or Effective Field Theory (EFT) approaches, to stabilize the transition~\cite{Efstratiou2025,Nojiri:2025uew,Silva:2025hxw,Scherer:2025esj}. Recent analyses incorporating DESI Data Release 2 (DR2)~\cite{DESI:2025zgx} combined with CMB observations favor a Dynamical Dark Energy scenario characterized by a specific phantom crossing behavior. Contrary to simple quintessence models, the data indicate a transition from a phantom-like phase ($w < -1$) in the past to a quintessence-like phase ($w > -1$) at late times, crossing the phantom divide at approximately $z \approx 0.4 - 0.5$, with a present-day value of $w_0 \approx -0.8$. 

Crucially, regarding the Hubble tension, this result presents a paradox: due to the geometric anti-correlation between $w_0$ and $H_0$, a present-day value of $w_0 > -1$ implies a lower inferred value for the Hubble constant. Consequently, rather than bridging the gap towards the Distance Ladder value ($H_0 \approx 73$ km s$^{-1}$ Mpc$^{-1}$), this specific form of dynamical dark energy risks exacerbating the discrepancy with the high Distance Ladder value. ~\cite{Pedrotti2025,Efstratiou:2025xou,Mirpoorian2025,Capozziello:2025qmh,Li:2025muv,Colgain:2025nzf,RoyChoudhury:2025dhe,Silva:2025hxw,Berti:2025phi,Jia:2025poj}. However, it aligns remarkably well with intermediate values: recent work by Zhang et al.\ using an interacting dark energy model finds $H_0 = 70.00 \pm 1.50\,\mathrm{km\,s^{-1}\,Mpc^{-1}}$ from CMB$+$DESI DR2 data (without SH0ES priors), a value consistent with intermediate estimates between Planck and SH0ES~\cite{zhang2025all}. If the DESI evidence for dynamical dark energy is confirmed, the observed shift between $\Lambda$CDM-dependent (Category 2) and model-independent (Category 3) measurements transforms from a statistical curiosity into a physical necessity. In this scenario, the Pure Local measurements (Category 3), yielding $H_0 \approx 71$ km s$^{-1}$ Mpc$^{-1}$, would emerge as the only methodologically sound Distance Ladder Independent/Sound Horizon Free determinations. Conversely, the lower values found in Category 2 would be interpreted as an artifact of imposing an invalid $\Lambda$CDM expansion history on local data. \cite{Efstratiou2025,Ozulker:2025ehg,Scherer:2025esj,Alestas2021,Silva:2025twg,Gialamas:2025pwv,Nojiri:2025uew,Roy:2025cxk,Santos:2025wiv,Abchouyeh:2025ans}.

\textbf{Ultralate-time models} propose environmental or physical changes at very low redshifts $z \lesssim 0.01$ that could affect the calibration of distance ladder indicators. The G-step model \cite{Marra2021,Ruchika2025,Perivolaropoulos:2025gzo,Perivolaropoulos:2022vql}, which posits a $\sim 4\%$ transition in the effective gravitational constant $G_{{eff}}$ at $z \approx 0.01$, represents a well-studied example. Related approaches involving screened fifth forces have been explored by Desmond et al.\ \cite{Desmond:2019ygn,Desmond:2020wep,Hogas:2025mii}, who showed that such modifications could systematically affect Cepheid and TRGB distance indicators differently. Alternative explanations within this category include local void scenarios, which posit that our local environment is underdense relative to the cosmic average, leading to an apparent acceleration of the local expansion rate~\cite{Kenworthy2019,Lukovic2020,Cai2021,Castello2022,Camarena2022,Marra2023,Jia2025,Banik:2025dlo}. While these models offer mechanisms to reconcile distance ladder and other measurements without modifying early-universe physics, they require fine-tuning and lack complete theoretical justification for the proposed transitions.

However, this ``Early vs.\ Late'' characterization may be oversimplified. It assumes all local measurements favor high $H_0$ values, an assumption challenged by Distance Ladder Independent/Sound Horizon Free measurements independent of the Cepheid/SN Ia calibration chain. Our recent analysis~\cite{Perivolaropoulos2024} demonstrated a stark bifurcation: Distance Ladder measurements clustered around $73\,\mathrm{km\,s^{-1}\,Mpc^{-1}}$, while independent methods statistically favored lower values closer to Planck. This suggests the discrepancy may lie between the distance ladder methodology and other probes, rather than between early and late Universe physics~\cite{Huang:2024}. Indeed, the tension has been shown to have a multidimensional character, with degeneracies between $H_0$, $\Omega_m$, and other parameters playing a crucial role~\cite{Pedrotti:2024kpn}.

In this work, we present an up-to-date compilation of 88 H0 measurements published through late-2025, all independent of the CMB Sound Horizon scale. We introduce a four-category classification based on methodology and model dependence: (1) Distance Ladder (30 measurements), (2) Local $\Lambda$CDM (33 measurements), (3) Pure Local (cosmological-model-independent, (16 measurements), and (4) CMB Sound Horizon Free (CMB--SHF) (9 measurements). The detailed definitions and classification criteria are presented in Section~\ref{subsec:classification}.

The classification follows a systematic decision process (Figure~\ref{fig:decision_tree}): measurements using multi-rung calibration chains are assigned to Category~1; those utilizing CMB data without $r_s$ to Category~4; remaining measurements are assigned to Category~2 or~3 based on whether they assume $\Lambda$CDM. The same observational technique can yield measurements in different categories depending on analysis methodology---for example, Megamaser observations with Keplerian models belong to Category~3~\cite{Kuo2013}, while those adopting $\Lambda$CDM priors belong to Category~2~\cite{Barua2025}.

This classification isolates two distinct tensions: the \emph{external tension} between the Distance Ladder and all other methods, and an \emph{internal tension} within these measurements examining whether $\Lambda$CDM assumptions systematically shift inferred $H_0$ values. A comprehensive overview of observational tensions and potential resolutions involving both systematics and fundamental physics has been presented in the CosmoVerse White Paper~\cite{CosmoVerseNetwork:2025alb}.

This paper is structured as follows: Section~\ref{sec:data_classification} details our data compilation and measurement tables. Section~\ref{sec:results} presents weighted means, reduced chi-squared values, and tension metrics. Section~\ref{sec:model_dependence} discusses model assumption impacts and bimodality in the $H_0$ distribution. Section~\ref{sec:conclusions} synthesizes these findings and their implications for the Distance Ladder Crisis.
%======================================================================
%\section{Data Compilation and Categorization}
%\label{sec:data_classification}
%======================================================================
% Goals for this section:
% 1. Briefly describe data selection criteria (sound horizon free, peer reviewed/arXiv).
% 2. Define the Four Categories in detail:
%    A. Distance Ladder (3-rung).
%    B. Local One-Step (LCDM dependent).
%    C. Local One-Step (Model Independent).
%    D. CMB Sound Horizon Free (LCDM dependent + CMB data).
% 3. Present the data tables (condensed or referenced) sorted by these categories.

%\subsection{Category 1: Distance Ladder Measurements}
% Brief description of the 30 measurements (Cepheids, TRGB, etc).

%\subsection{Category 2: Local ($\Lambda$CDM assumption)}
% Description of measurements like Time Delays or Sirens that assume LCDM shape.

%\subsection{Category 3: Pure Local (Model Independent)}
% Description of measurements like Cosmic Chronometers or geometric methods without shape assumptions.

%\subsection{Category 4: CMB Sound Horizon Free}
% Description of measurements utilizing CMB data but not r_s (e.g., early ISW, horizon scale equality).

%======================================================================
\section{Data Compilation and Categorization}
\label{sec:data_classification}
%======================================================================

To dissect the Hubble tension beyond the traditional ``Early vs.\ Late'' dichotomy, we have compiled a comprehensive set of 88 independent measurements of the Hubble constant, updated through late-2025. This compilation builds upon the catalog presented in our previous work~\cite{Perivolaropoulos2024}, significantly expanded to include recent results from James Webb Space Telescope (JWST) calibration campaigns, the third generation of gravitational wave catalogs (GWTC-3), and the latest releases from the Dark Energy Spectroscopic Instrument (DESI)~\cite{DESI:2025zgx}.

\subsection{Selection Criteria}
\label{subsec:selection}

Our selection criteria are strictly defined to isolate the impact of the Sound Horizon scale ($r_s$) and the Distance Ladder methodology:
\begin{enumerate}
    \item \textbf{Sound Horizon Independence:} We rigorously exclude any measurement that uses the CMB Sound Horizon at the drag epoch as a standard ruler. This excludes standard Baryon Acoustic Oscillation (BAO) inverse Distance Ladder constraints and the standard Planck $\Lambda$CDM inference.
    \item \textbf{Public Availability:} All measurements are sourced from peer-reviewed journals or preprints available on arXiv.
    \item \textbf{Independence of Estimates:} Where multiple analyses utilize the exact same sample (e.g., multiple re-analyses of the same supernova catalog with identical calibration), we select the most recent or representative result to avoid double-counting, unless the methodologies differ significantly in their systematic assumptions.
    \item \textbf{Methodological Transparency:} We include only measurements where the original publication clearly states the cosmological assumptions employed, allowing unambiguous category assignment.
\end{enumerate}

\subsection{Classification Methodology}
\label{subsec:classification}

The compiled measurements are classified into four distinct categories based on two primary criteria: (i) reliance on the local Distance Ladder calibration chain, and (ii) dependence on the assumed cosmological model, specifically the shape of the expansion history $H(z)$. The classification follows the decision process outlined in Section~\ref{sec:introduction} and is determined by the explicit methodological statements in each original publication.

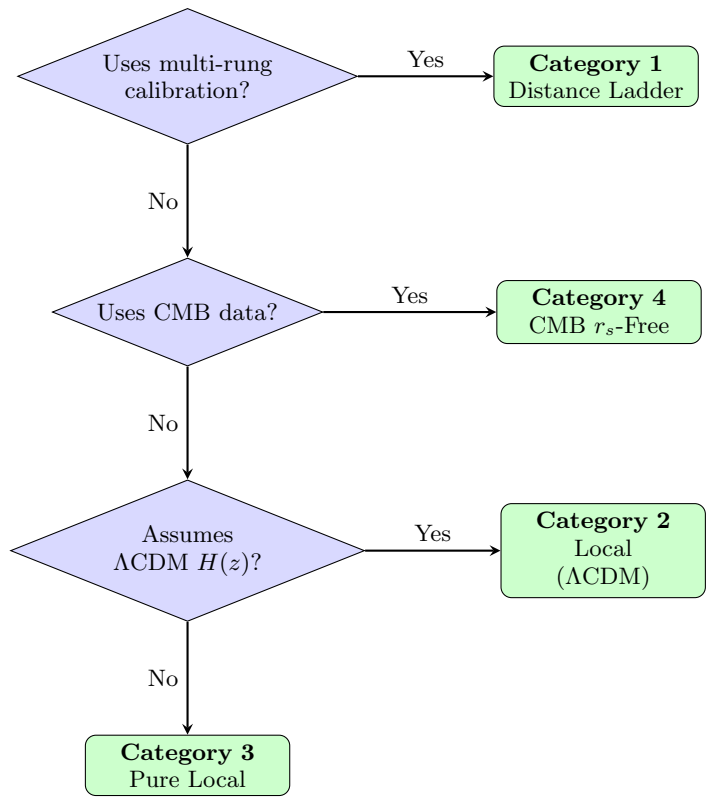
\begin{figure}[h]
\centering
\begin{tikzpicture}[
    node distance=1.5cm,
    decision/.style={diamond, draw, fill=blue!15, text width=2.8cm, text centered, inner sep=1pt, aspect=2.5},
    block/.style={rectangle, draw, fill=green!20, text width=2.5cm, text centered, rounded corners, minimum height=0.8cm},
    arrow/.style={thick,->,>=stealth}
]
% Nodes
\node[decision] (ladder) {Uses multi-rung calibration?};
\node[block, right=1.8cm of ladder] (cat1) {\textbf{Category 1}\\Distance Ladder};
\node[decision, below=1.5cm of ladder] (cmb) {Uses CMB data?};
\node[block, right=2.3cm of cmb] (cat4) {\textbf{Category 4}\\CMB $r_s$-Free};
\node[decision, below=1.5cm of cmb] (lcdm) {Assumes $\Lambda$CDM $H(z)$?};
\node[block, right=1.8cm of lcdm] (cat2) {\textbf{Category 2}\\Local \\($\Lambda$CDM)};
\node[block, below=1.5cm of lcdm] (cat3) {\textbf{Category 3}\\Pure Local\\};

% Arrows
\draw[arrow] (ladder) -- node[above] {Yes} (cat1);
\draw[arrow] (ladder) -- node[left] {No} (cmb);
\draw[arrow] (cmb) -- node[above] {Yes} (cat4);
\draw[arrow] (cmb) -- node[left] {No} (lcdm);
\draw[arrow] (lcdm) -- node[above] {Yes} (cat2);
\draw[arrow] (lcdm) -- node[left] {No} (cat3);
\end{tikzpicture}
\caption{Decision tree for classifying \texorpdfstring{$H_0$}{H0} measurements into the four categories. Category assignments are based on explicit methodological statements in the original publications.}
\label{fig:decision_tree}
\end{figure}

Figure~\ref{fig:decision_tree} illustrates the classification decision tree schematically. For each measurement, we first assess whether it employs a multi-rung calibration chain (leading to Category~1). If not, we determine whether it utilizes CMB data (leading to Category~4 if Sound-Horizon-Free). For the remaining local measurements, we examine whether the analysis assumes a $\Lambda$CDM expansion history (Category~2) or proceeds cosmological-model-independently (Category~3).

It is important to emphasize that the same observational technique can yield measurements in different categories depending on the analysis approach adopted by the authors. Table~\ref{tab:technique_categories} provides illustrative examples of how specific techniques are distributed across categories based on their methodological implementation.

Throughout this work, we use the term ``cosmological-model-independent'' (Category~3) to denote measurements that do not assume a specific functional form for the expansion history $H(z)$---in particular, they do not assume the standard $\Lambda$CDM expansion history. This should be understood as 
independence from cosmological model assumptions rather than complete model independence, as all measurements necessarily rely on certain physical assumptions (e.g., standard astrophysics, the FLRW metric, general relativity). Similarly, we define Distance Ladder-Independent/Sound Horizon Free (DLI--SHF) measurements (Categories~2, 3, and~4) as any $H_0$ determination that bypasses the traditional three-rung Distance Ladder calibration chain (geometric anchors $\to$ Cepheids/TRGB/JAGB/Miras $\to$ SNe~Ia in the Hubble flow), regardless of whether it probes early or late cosmic times. Some DLI--SHF methods may involve their own calibration procedures (e.g., using cosmic chronometers to calibrate the HII galaxy Hubble diagram), but these do not rely on the standard candle--supernova chain that defines the Distance Ladder.

\begin{table}[h]
\caption{Examples of how the same observational technique can be assigned to different categories based on analysis methodology.}
\label{tab:technique_categories}
\scriptsize
\setlength{\tabcolsep}{3pt}
\begin{tabular}{@{}p{2.2cm}|p{2.3cm}|p{2.5cm}@{}}
\toprule
Technique & Category 2 ($\Lambda$CDM) & Category 3 (Cosmological-Model-Independent) \\
\midrule
Megamasers & Barua (2025): uses $\Lambda$CDM $D_{A}(z)$ & Kuo (2013): Keplerian disk only \\
\addlinespace
Strong Lensing & Shajib (2023): $\Lambda$CDM lens model & Du (2023): cosmographic approach \\
\addlinespace
Gravitational Waves & Abbott (2021): $\Lambda$CDM $d_L(z)$ & Sneppen (2023): EPM + kilonova \\
\addlinespace
Cosmic Chronom.& Moresco (2023): $\Lambda CDM$ fit $H(z)$  & Zhang (2022): Gaussian Process  \\
\bottomrule
\end{tabular}
\end{table}

We now describe each category in detail and present the compiled measurements.

%---------------------------------------------------------------------
\subsection{Category 1: Distance Ladder Measurements}
\label{subsec:cat1_ladder}
%---------------------------------------------------------------------

This category comprises 30 measurements that employ the traditional multi-rung calibration method. These measurements anchor the absolute magnitude of Type~Ia supernovae (SNe~Ia) in the Hubble flow using local standard candles calibrated by geometric distances. This is the ``Late Universe'' standard, relying on the fundamental assumption that the physics of calibrators (Cepheids, TRGB, JAGB, Miras) and SNe~Ia are uniform across cosmic time and environment.

The three-rung structure proceeds as follows: (1)~geometric anchors (Gaia parallaxes, NGC~4258 Megamaser distance, LMC eclipsing binaries) establish absolute distances to nearby calibrators; (2)~these calibrators (Cepheids, TRGB, JAGB, Miras) are observed in galaxies that also host SNe~Ia, transferring the distance scale; (3)~the calibrated SNe~Ia absolute magnitude is applied to supernovae in the Hubble flow ($0.023 < z < 0.15$) to infer $H_0$.

Our compilation (Table~\ref{tab:cat1_ladder}) includes Cepheid-based measurements and recent JWST updates~\cite{Riess2025}, TRGB-based measurements including CCHP results~\cite{Freedman2024}, and alternative calibrators such as JAGB stars, Mira variables, Surface Brightness Fluctuations (SBF), and the Tully-Fisher relation. We also include the galaxy pairwise velocity method~\cite{Zhang2024}, which, while methodologically distinct, relies on calibration from distance ladder measurements~\cite{Tsagas:2025pxi,Duangchan:2025uzj,Salzano:2025ang}.

\begin{table}[h]
\caption{\textbf{Category 1: Distance Ladder Measurements.} 13 selected representative entries from the 30 measurements in this category. The full table is provided in Appendix~\ref{app:tables}.}
\label{tab:cat1_ladder}
\scriptsize
\setlength{\tabcolsep}{2pt}
\begin{tabular}{@{}llcc l@{}}
\toprule
Author & Year & $H_0$ & $\sigma$ & Method \\
\midrule
Huang~\cite{Huang2020} & 2019 & 73.3 & 4.0 & Mira \\
de Jaeger~\cite{deJaeger2020} & 2020 & 75.8 & 5.1 & Cepheids/TRGB + SNe II \\
Khetan~\cite{Khetan2021} & 2021 & 70.5 & 4.1 & SBF \\
Freedman~\cite{Freedman2021} & 2021 & 69.8 & 1.7 & TRGB (CCHP) \\
Kenworthy~\cite{Kenworthy2022} & 2022 & 73.1 & 2.5 & Two-rung ladder \\
Scolnic~\cite{Scolnic2024} & 2023 & 73.22 & 2.06 & TRGB (SH0ES calibrators) \\
Li~\cite{Li2024b} & 2024 & 74.7 & 3.1 & JAGB + SNe Ia \\
Lee~\cite{Lee2024} & 2024 & 67.8 & 2.7 & JAGB + SNe Ia \\
Lapuente~\cite{lapuente2025} & 2025 & 72.61 & 1.69 & TRGB+JAGB+Ceph.+SNe Ia (twins)\\
Jensen~\cite{Jensen2025} & 2025 & 73.8 & 2.4 & TRGB \\
Riess~\cite{Riess2025} & 2025 & 73.49 & 0.93 & Cepheids (JWST) \\
Bhardwaj~\cite{Bhardwaj2025} & 2025 & 73.06 & 2.6 & Mira \\
Newman~\cite{Newman2025} & 2025 & 75.3 & 2.9 & TRGB + SNe Ia \\
\bottomrule
\end{tabular}
\end{table}

We note that the Lee~\cite{Lee2024} JAGB measurement ($H_0 = 67.8 \pm 2.7$~km~s$^{-1}$~Mpc$^{-1}$) represents a notable outlier within this category, yielding a value consistent with Planck rather than the typical Distance Ladder result. This discrepancy with other JAGB analyses (e.g., Li, Lapuente~\cite{Li2024b,lapuente2025}) appears to originate from differences in the JAGB zero-point calibration and photometric selection criteria, highlighting that even within a single indicator class, methodological choices can significantly impact the inferred $H_0$. A critical recent addition to this category is the analysis by Ruiz-Lapuente et al.~\cite{lapuente2025}, which utilizes the ``SNe Ia twins'' method calibrated via both TRGB and JAGB. Notably, their specific calibration using JAGB yields $H_0 = 72.34$ km s$^{-1}$ Mpc$^{-1}$, while their TRGB determination gives $72.54$ km s$^{-1}$ Mpc$^{-1}$. In our compilation (Table~\ref{tab:cat1_ladder}), we have adopted the value $H_0 = 72.61 \pm 1.69$ km s$^{-1}$ Mpc$^{-1}$, which represents the weighted average across the different methodological combinations presented in their work. This independent confirmation of high $H_0$ values using alternative calibrators is particularly significant as it directly challenges previous lower estimates ($H_0 \sim 69.8$ km s$^{-1}$ Mpc$^{-1}$) historically associated with the CCHP program. It suggests that the previously reported discrepancies between Cepheids and other local calibrators of the Distance Ladder may have been overstated, implying that unrecognized systematics in the specific calibration or sample selection of earlier CCHP analyses—rather than the indicators themselves—may account for the lower values. Consequently, this result reinforces the internal consistency of Category 1, further isolating the Distance Ladder methodology from the rest of the cosmological probes.

%---------------------------------------------------------------------
\subsection{Category 2: Local (\texorpdfstring{$\Lambda$}{Λ}CDM Assumption)}
\label{subsec:cat2_lcdm}
%---------------------------------------------------------------------

This category includes 33 measurements that bypass the Distance Ladder but infer $H_0$ by explicitly assuming the standard $\Lambda$CDM expansion history:
\begin{equation}
H(z) = H_0 \sqrt{\Omega_m (1+z)^3 + 1 - \Omega_m}.
\label{eq:lcdm_hz}
\end{equation}
While these methods probe the local Universe (typically $z < 1.5$), they rely on this model to map observables---such as time delays, luminosity distances, dispersion measures, or angular diameter distances---to $H_0$. The key criterion for inclusion in this category is that the original analysis \emph{explicitly assumes} the functional form of Eq.~\eqref{eq:lcdm_hz}, often with $\Omega_m$ fixed to a fiducial value (typically $\Omega_m \approx 0.3$) or marginalized over a narrow prior.

This category includes:
\begin{itemize}
    \item \textbf{Gravitational Wave Standard Sirens:} Both bright sirens (with EM counterparts, e.g., GW170817~\cite{Palmese2023}) and dark sirens (using galaxy catalogs for redshift inference~\cite{Abbott2021,Beirnaert2025}), where the luminosity distance $d_L(z)$ is computed assuming $\Lambda$CDM.
    \item \textbf{Strong Lensing Time Delays:} Analyses where the cosmological model enters both the time-delay distance calculation and often the lens mass modeling~\cite{Shajib2023, Kelly2023, Birrer2025}.
    \item \textbf{Fast Radio Bursts:} Measurements using the dispersion measure--redshift relation interpreted under $\Lambda$CDM assumptions for the intergalactic medium~\cite{Wu2022, PiratovMoreno2025}.
\end{itemize}

\begin{table}[h]
\caption{\textbf{Category 2: Local ($\Lambda$CDM Assumption).} 12 selected representative entries from the 33 measurements in this category. The full table is provided in Appendix~\ref{app:tables}}
\label{tab:cat2_lcdm}
\scriptsize
\setlength{\tabcolsep}{2pt}
\begin{tabular}{@{}llcc l@{}}
\toprule
Author & Year & $H_0$ & $\sigma$ & Method \\
\midrule
Wu~\cite{Wu2022} & 2021 & 68.8 & 4.7 & FRBs \\
Shajib~\cite{Shajib2023} & 2023 & 77.1 & 7.2 & Lensing (TDCOSMO) \\
Napier~\cite{Napier2023} & 2023 & 74.1 & 8.0 & Cluster Lenses \\
Kelly~\cite{Kelly2023} & 2023 & 66.6 & 3.7 & Lensed SN Refsdal \\
Liu~\cite{liu2023h} & 2023 & 59.1 & 3.6 & Cluster Lensed Quasar \\
Martinez~\cite{Martinez2023} & 2023 & 74.0 & 11.2 & Time-Delay Cosmography \\
Pascale~\cite{Pascale2024} & 2024 & 71.8 & 8.7 & Lensed SN H0pe (JWST) \\
Yang~\cite{Yang2024} & 2024 & 74.0 & 7.4 & FRBs w/ Scattering \\
Piratova~\cite{PiratovMoreno2025} & 2025 & 65.1 & 7.4 & FRBs (MLE) \\
Beirnaert~\cite{Beirnaert2025} & 2025 & 79.0 & 9.0 & GW Dark Sirens \\
Birrer~\cite{Birrer2025} & 2025 & 71.6 & 3.6 & TDCOSMO Strong Lensing \\
Pierel~\cite{Pierel2025,Suyu:2025hjn} & 2025 & 66.9 & 9.8 & SN Encore Time Delays \\
\bottomrule
\end{tabular}
\end{table}

%---------------------------------------------------------------------
\subsection{Category 3: Pure Local (Cosmological-Model Independent)}
\label{subsec:cat3_pure}
%---------------------------------------------------------------------

This category represents the cleanest test of the Distance Ladder, as it comprises 16 measurements that determine $H_0$ geometrically or via direct measurement of the expansion rate, \emph{without} assuming a specific functional form for $H(z)$. These measurements are either:
\begin{itemize}
    \item \textbf{Purely geometric:} Relying on Keplerian dynamics or light-travel-time arguments that are independent of the global cosmological model (e.g., Megamaser disk modeling~\cite{Kuo2013,Gao2015}).
   \item \textbf{Direct $H(z)$ measurements:} Cosmic chronometers measure $H(z) = -\frac{1}{1+z}\frac{dz}{dt}$ directly from the differential aging 
of massive, passively evolving galaxies, providing cosmological-model-independent determinations of the expansion rate at various redshifts~\cite{Moresco2023}. 
To extract $H_0$, two main approaches have been employed: (i) Gaussian Process reconstruction of $H(z)$ combined with a cosmographic expansion of the 
luminosity distance in the Hubble flow ($0.02 < z < 0.15$), (ii) using cosmic chronometers to calibrate the H\,\textsc{ii} galaxies Hubble diagram, which in turn anchors Type Ia supernovae at low redshift via Hubble's law~\cite{Zhang2022}. Both methods avoid assumptions about the 
underlying cosmological model beyond the FLRW metric.
    \item \textbf{Cosmographic approaches:} Analyses that expand $H(z)$ or $d_L(z)$ in Taylor series around $z=0$, avoiding assumptions about the energy content of the Universe~\cite{Du2023}.
    \item \textbf{Expanding photosphere methods:} Geometric distance determinations from supernovae using photospheric velocity and angular size evolution~\cite{Sneppen2023, Vogl2024}.
\end{itemize}

The key distinction from Category~2 is that these analyses do \emph{not} assume the validity of Eq.~\eqref{eq:lcdm_hz}. For Megamaser measurements, we include in this category only those analyses that rely purely on Keplerian disk modeling without adopting $\Lambda$CDM-based peculiar velocity corrections or external distance priors tied to the Distance Ladder.

\begin{table}[h]
\caption{\textbf{Category 3: Pure Local (Cosmological-Model Independent).} 9 representative measurements in this category, representing the cosmological-model-independent local determinations of $H_0$. The full table is provided in Appendix~\ref{app:tables}}
\label{tab:cat3_pure}
\scriptsize
\setlength{\tabcolsep}{2pt}
\begin{tabular}{@{}llcc l@{}}
\toprule
Author & Year & $H_0$ & $\sigma$ & Method \\
\midrule
Kuo~\cite{Kuo2013} & 2013 & 68.0 & 9.0 & Megamasers (MCP) \\
Gao~\cite{Gao2015} & 2015 & 66.0 & 6.0 & Megamasers \\
Zhang~\cite{Zhang2022} & 2022 & 65.9 & 3.0 & CC + HII \\
Du~\cite{Du2023} & 2023 & 71.5 & 3.8 & TD Lenses + GRBs \\
Sneppen~\cite{Sneppen2023} & 2023 & 67.0 & 3.6 & EPM + Kilonovae \\
Li~\cite{Li2024a} & 2024 & 66.3 & 3.7 & Lensing Bias Correction \\
Vogl~\cite{Vogl2024} & 2024 & 74.9 & 2.7 & SNe II (Tailored EPM) \\
Song~\cite{Song2025} & 2025 & 70.4 & 6.9 & GW + SGL Calibration \\
Favale~\cite{Favale2025} & 2025 & 68.8 & 3.0 & Cosmic Chronometers \\
\bottomrule
\end{tabular}
\end{table}

We highlight that this category exhibits significant internal heterogeneity. Notably, the Vogl~\cite{Vogl2024} tailored EPM measurement ($H_0 = 74.9 \pm 2.7$~km~s$^{-1}$~Mpc$^{-1}$) strongly favor the Distance Ladder value, while most cosmic chronometer measurements favor lower values. This heterogeneity will be discussed further in Section~\ref{sec:model_dependence}.

%---------------------------------------------------------------------
\subsection{Category 4: CMB Sound-Horizon-Free (CMB--SHF)}
\label{subsec:cat4_cmb_free}
%---------------------------------------------------------------------

The final category includes 9 measurements that utilize early-Universe data (primarily CMB) but specifically avoid the Sound Horizon $r_s$ as a standard ruler. Instead, these methods rely on alternative physical scales that are calibrated independently of pre-recombination acoustic physics:

\begin{itemize}
    \item \textbf{Matter-radiation equality scale ($k_{\rm eq}$):} The comoving wavenumber corresponding to the horizon size at matter-radiation equality provides a standard ruler that depends on $\Omega_m h^2$ and $\Omega_r h^2$ but is independent of $r_s$~\cite{Philcox2022, Bahr-Kalus2025}.
    \item \textbf{BAO without $r_s$ priors:} Analyses that use BAO peak positions but treat $r_s$ as a free parameter, constraining $H_0$ through the shape of $H(z)$ rather than absolute calibration~\cite{Pogosian2024}.
    \item \textbf{Sunyaev-Zel'dovich (SZ) effect:} Combining SZ decrements with X-ray observations of galaxy clusters to infer angular diameter distances independent of $r_s$~\cite{Reese2003, Colaco2023}.
\end{itemize}

\begin{table}[h]
\begin{threeparttable}
\caption{\textbf{Category 4: CMB Sound Horizon Free.} All 9 measurements in this category, utilizing early-Universe data without the Sound Horizon scale.}
\label{tab:cat4_cmb_free}
\scriptsize
\setlength{\tabcolsep}{2pt}
\begin{tabular}{@{}llcc l@{}}
\toprule
Author & Year & $H_0$ & $\sigma$ & Method \\
\midrule
Reese~\cite{Reese2003} & 2003 & 61.0 & 18.0 & SZ Effect \\
Philcox~\cite{Philcox2022}& 2022 & 64.8 & $^{+2.2}_{-2.5}$ & BAO scale,Matter--rad.\ eq.\\
Colaco~\cite{Colaco2023} & 2023 & 67.2 & 6.1 & SZ + X-ray \\
Pogosian~\cite{Pogosian2024} & 2024 & 68.05 & 0.94 & BAO-pre DESI (No $r_s$) \\
Pogosian\tnote{a}~\cite{Pogosian2025priv} & 2025 & 69.37 & 0.65 & BAO-DESI DR2 (No $r_s$) \\
Bahr-Kalus~\cite{Bahr-Kalus2025} & 2025 & 65.2 & 5.6 & DESI Turnover Scale \\
Garcia~\cite{GarciaEscudero:2025}& 2025 & 70.03 & 0.97 & BAO-DESI DR2 +$\theta^*+A_s$ (No $r_s$) \\
Zaborowski~\cite{Zaborowski2025} & 2025 & 70.8 & $^{+2.0}_{-2.2}$ & BAO-DESI +$\theta^*$ (No $r_s$)\\
Krolewski~\cite{Krolewski2025} & 2025 & 69.0 & 2.5 & DESI + CMB (No $r_s$) \\ 
\bottomrule
\end{tabular}
\begin{tablenotes}[flushleft]
\footnotesize
\item[a] Private communication.
\end{tablenotes}
\end{threeparttable}
\end{table}

We note that this category includes two entries from Pogosian et al.~\cite{Pogosian2024,Pogosian2025priv}, corresponding to analyses using DESI and pre-DESI BAO data respectively. While these share a common methodology, they utilize substantially different datasets and thus provide partially independent constraints. 

\subsection{Summary of Classification}
\label{subsec:classification_summary}

Table~\ref{tab:category_summary} summarizes the four categories and their key characteristics. The complete measurement tables with extended methodological information are provided in Appendix~\ref{app:tables}.

\begin{table}[h]
\caption{Summary of the four measurement categories.}
\label{tab:category_summary}
\scriptsize
\setlength{\tabcolsep}{3pt}
\begin{tabular}{@{}clccc@{}}
\toprule
Cat. & Description & $N$ & Uses Ladder? & Assumes $\Lambda$CDM? \\
\midrule
1 & Distance Ladder & 30 & Yes & No \\
2 & Local ($\Lambda$CDM) & 33 & No & Yes \\
3 & Pure Local & 16 & No & No \\
4 & CMB--SHF & 9 & No & Yes$^*$ \\
\bottomrule
\end{tabular}
\\ \scriptsize{$^*$Category 4 assumes $\Lambda$CDM for $H(z)$ shape but avoids the Sound Horizon $r_s$.}
\end{table}

%======================================================================
%\section{Statistical Analysis and Results}
%\label{sec:results}
%======================================================================
% Goals for this section:
% 1. Present the Weighted Mean and Chi-squared for each of the 4 categories.
% 2. Highlight the main values:
%    - Ladder: High (~73)
%    - LCDM: Lowest (~67)
%    - Pure: Intermediate (~70)
%    - CMB No-rs: Low (~68.7)
% 3. Visual comparison (Plot of the 4 means vs Planck).

%\subsection{Weighted Mean Statistics by Category}
% The numerical results.

%\subsection{Quantifying the Tension}
% Calculate sigma differences between Category 1 vs the others.

%======================================================================
\section{Statistical Analysis and Results}
\label{sec:results}
%======================================================================

In this section, we present a detailed statistical analysis of the $H_0$ measurements classified in Section~\ref{sec:data_classification}. We utilize the inverse-variance weighted mean to determine the central value for each category and the reduced chi-squared statistic ($\chi^2_\nu$) to assess internal consistency. We explicitly quantify the tension between the Distance Ladder (Category~1) and the various subsets of Sound-Horizon-Free measurements, concluding with a comparison against the combined ensemble of all Distance Ladder Independent/Sound Horizon Free (DLI--SHF) methods. We also compare our results with our previous analysis~\cite{Perivolaropoulos2024} to assess the evolution of the tension landscape.

\subsection{Methodology}
\label{subsec:methodology}

For a set of $N$ independent measurements $\{H_{0,i} \pm \sigma_i\}$, the weighted mean $\hat{H}_0$ and its standard error $\sigma_{\hat{H}_0}$ are calculated as:
\begin{equation}
\hat{H}_0 = \frac{\sum_{i=1}^{N} w_i H_{0,i}}{\sum_{i=1}^{N} w_i}, \quad \sigma_{\hat{H}_0} = \left( \sum_{i=1}^{N} w_i \right)^{-1/2},
\label{eq:weighted_mean}
\end{equation}
where the weights are given by $w_i = 1/\sigma_i^2$. This estimator minimizes the variance for Gaussian-distributed errors. 

To evaluate whether the variation within a category is consistent with the reported uncertainties, we compute the reduced chi-squared:
\begin{equation}
\chi^2_\nu = \frac{1}{N-1} \sum_{i=1}^{N} \frac{(H_{0,i} - \hat{H}_0)^2}{\sigma_i^2}.
\label{eq:chi2}
\end{equation}

In cases of asymmetric errors, the following Equation is adopted for computations involving the mean or a unified error value:
\begin{equation}
\sigma = \sqrt{\frac{(\sigma^{+})^2 + (\sigma^{-})^2}{2}}
\end{equation}

The interpretation of $\chi^2_\nu$ provides important diagnostic information:
\begin{itemize}
    \item $\chi^2_\nu \approx 1$: Good internal consistency; the scatter is consistent with reported uncertainties.
    \item $\chi^2_\nu \gg 1$: Underestimated errors, unrecognized systematics, or genuine physical scatter.
    \item $\chi^2_\nu \ll 1$: Overestimated errors, unaccounted correlations between measurements, or possible publication bias toward a consensus value.
\end{itemize}

The statistical tension between two independent estimates $A$ and $B$ is quantified in terms of standard deviations ($n_\sigma$) as:
\begin{equation}
n_\sigma = \frac{|\hat{H}_{0,A} - \hat{H}_{0,B}|}{\sqrt{\sigma_A^2 + \sigma_B^2}}.
\label{eq:tension_metric}
\end{equation}

We note that this tension metric assumes statistical independence between the two samples. In practice, correlations exist within each category due to shared calibration data, common analysis frameworks, and overlapping samples. We address the impact of these correlations in Appendix~\ref{app:correlations}, where we show that accounting for realistic correlation structures reduces the tension significance but does not qualitatively alter our conclusions.

\subsection{Weighted Mean Statistics by Category}
\label{subsec:weighted_means}

\begin{figure*}[t]
    \centering
    \includegraphics[width=\textwidth]{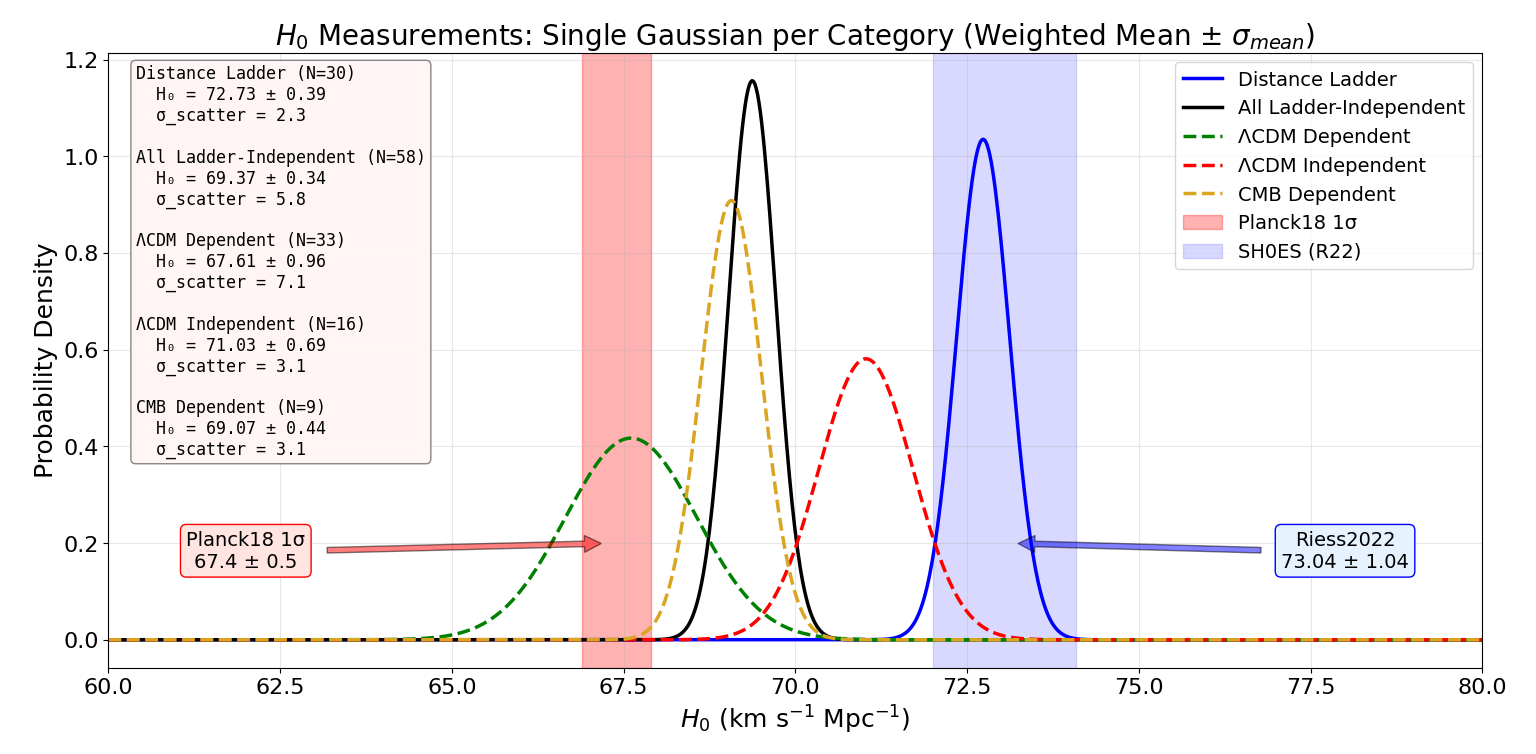}
    \caption{\textbf{Gaussian probability distributions of the four measurement categories.} The visual separation is striking: Category~1 (Distance Ladder, blue peak) is isolated at high $H_0$. Categories~2 (Local $\Lambda$CDM) and 4 (CMB--SHF) overlap significantly at low $H_0$, demonstrating that local data analyzed under $\Lambda$CDM align with early-universe constraints. Category~3 (Pure Local) occupies a broader, intermediate region, highlighting the shift in central value when $\Lambda$CDM model assumptions are relaxed. The black curve represents the ensemble of all DLI--SHF measurements. The Planck 2018 value ($H_0 = 67.4 \pm 0.5$~km~s$^{-1}$~Mpc$^{-1}$) and SH0ES value ($H_0 = 73.04 \pm 1.04$~km~s$^{-1}$~Mpc$^{-1}$) are indicated for reference.}
    \label{fig:Gaussian_probability_distributions}
\end{figure*}

The statistical results for the four categories defined in Section~\ref{sec:data_classification} are summarized in Table~\ref{tab:results_summary}. For comparison, we also include the results from our previous analysis~\cite{Perivolaropoulos2024}, which employed a binary classification scheme (Distance Ladder vs.\ Distnance Ladder Independent/Sound Horizon Free).

\begin{table}[ht]
\caption{\textbf{Statistical Summary of Sound-Horizon-Free Categories.} The weighted mean $H_0$ (km~s$^{-1}$~Mpc$^{-1}$), uncertainty, reduced chi-squared ($\chi^2_\nu$), and tension relative to Category~1 (Distance Ladder). The bottom section shows the combined DLI--SHF results and, for comparison, the results from our 2024 analysis~\cite{Perivolaropoulos2024}.}
\label{tab:results_summary}
\setlength{\tabcolsep}{3pt}
\footnotesize
\begin{tabular}{@{}lcccc@{}}
\toprule
Category & $N$ & $H_0 \pm \sigma$ & $\chi^2_\nu$ & Tension \\
\midrule
\textbf{1. Distance Ladder} & 30 & $72.73 \pm 0.39$ & 0.72 & --- \\
\textbf{2. Local ($\Lambda$CDM)} & 33 & $67.61 \pm 0.96$ & 0.80 & $4.9\sigma$ \\
\textbf{3. Pure Local} & 16 & $71.03 \pm 0.69$ & 0.77 & $2.2\sigma$ \\
\textbf{4. CMB--SHF} & 9 & $69.07 \pm 0.44$ & 0.89 & $6.1\sigma$ \\
\midrule
\textit{Combined (2+3+4)} & 58 & $69.37 \pm 0.34$ & 0.95 & $6.5\sigma$ \\
\textit{Corr.-adjusted (2+3+4)} & 58 & $69.52 \pm 0.42$ & --- & $3.9\sigma$ \\
\midrule
\multicolumn{5}{l}{\textit{Comparison with Perivolaropoulos (2024)~\cite{Perivolaropoulos2024}:}} \\
\textit{2024 Distance Ladder} & 20 & $72.8 \pm 0.5$ & 0.51 & --- \\
\textit{2024 DLI--SHF} & 33 & $68.3 \pm 0.5$ & 0.95 & $\sim 5\sigma$ \\
\bottomrule
\end{tabular}
\end{table}

To visualize the distinct behaviors of these categories, we present the Gaussian probability distributions in Figure~\ref{fig:Gaussian_probability_distributions}. This Figure illustrates the probability density functions for each of the four subgroups, offering a clear visual representation of the tensions quantified in Table~\ref{tab:results_summary}.

\subsubsection{Category 1: Distance Ladder \texorpdfstring{($H_0 = 72.73 \pm 0.39~km~s^{-1}~Mpc^{-1}$)}{H0 = 72.73 +/- 0.39 km s\textasciicircum-1 Mpc\textasciicircum-1}}

The 30 measurements comprising the Distance Ladder category yield a weighted mean of $H_0 = 72.73 \pm 0.39$~km~s$^{-1}$~Mpc$^{-1}$. As shown in Figure~\ref{fig:Gaussian_probability_distributions}, this category forms a sharp, narrow peak that is clearly separated from the other distributions. 

The reduced chi-squared $\chi^2_\nu = 0.72$ indicates good internal consistency, with the scatter being slightly smaller than expected from the reported uncertainties. This value is notably higher than our 2024 analysis ($\chi^2_\nu = 0.51$), suggesting that the expanded data---which now includes measurements with greater methodological diversity (e.g., JAGB, Miras, pairwise velocities)---exhibits more realistic scatter. Nevertheless, $\chi^2_\nu < 1$ may indicate that: (i) reported uncertainties are somewhat conservative, (ii) positive correlations exist between measurements sharing common calibration anchors, or (iii) there is a tendency for published results to cluster around previously established values.

The high precision of this combined estimate is driven by the recent JWST Cepheid measurements~\cite{Riess2025} and compatible results from other calibrators including TRGB~\cite{Jensen2025}, Miras~\cite{Bhardwaj2025}, and SBF~\cite{Blakeslee2021}. We note that the Lee et al.~\cite{Lee2024} JAGB measurement ($H_0 = 67.8 \pm 2.7$) represents a $\sim 2\sigma$ outlier within this category, as discussed in Section~\ref{sec:data_classification}.

\subsubsection{Category 2: Local \texorpdfstring{$\Lambda$CDM}{LambdaCDM} \texorpdfstring{($H_0 = 67.61 \pm 0.96$~km~s$^{-1}$~Mpc$^{-1}$)}{(H0 = 67.61 +/- 0.96 km s-1 Mpc-1)}}

The 33 measurements in this category, which utilize local data but assume the $\Lambda$CDM expansion history, yield the \emph{lowest} aggregate value of $H_0 = 67.61 \pm 0.96$~km~s$^{-1}$~Mpc$^{-1}$. This value is fully consistent with, and in fact slightly lower than, the Planck CMB inference ($67.4 \pm 0.5$~km~s$^{-1}$~Mpc$^{-1}$), despite being derived entirely from local observations.

The reduced chi-squared $\chi^2_\nu = 0.80$ indicates internal consistency among these diverse methods (Strong Lensing, Gravitational Wave sirens, FRBs). This near-unity value suggests that the reported uncertainties accurately capture the true scatter and that no significant unrecognized systematics are present within this category. 

In Figure~\ref{fig:Gaussian_probability_distributions}, the distribution of category 2 appears at the lowest end of the parameter space, almost coincident with the Planck CMB value. This remarkable alignment between local $\Lambda$CDM-dependent measurements and early-universe physics---achieved without using the Sound Horizon---is one of the central findings of this work.

\begin{figure}[t]
    \centering
    \hspace{-1cm}
    \includegraphics[width=\columnwidth, height=8cm]{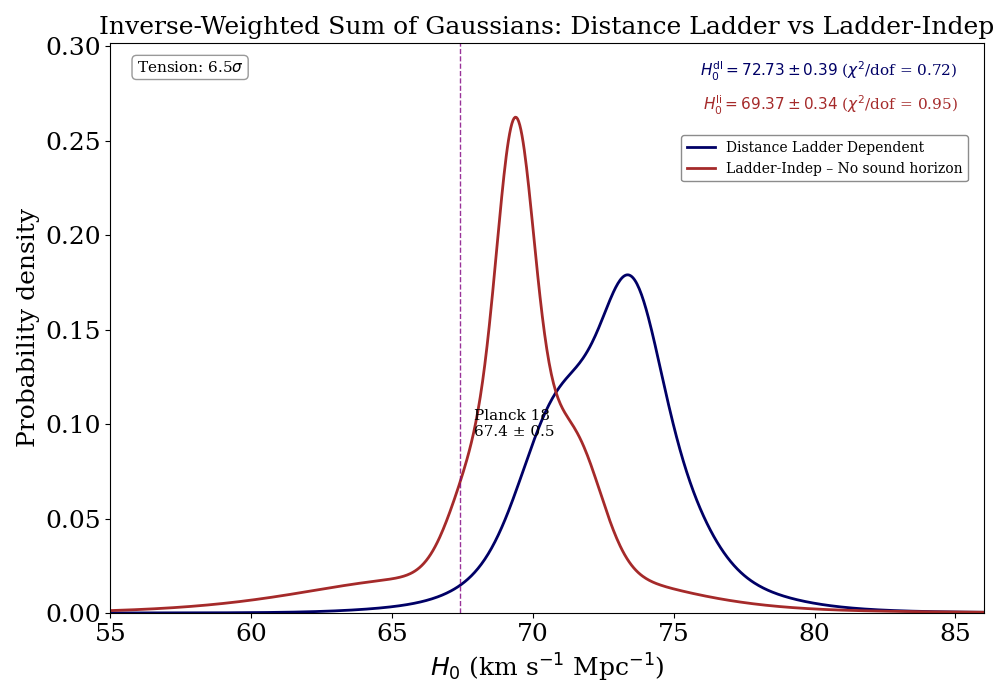}
    \caption{The blue curve represents Category~1 (Distance Ladder, $H_0 = 72.73 \pm 0.39$~km~s$^{-1}$~Mpc$^{-1}$), while the red curve represents the combined ensemble of all DLI--SHF measurements (Categories~2+3+4, $H_0 = 69.37 \pm 0.34$~km~s$^{-1}$~Mpc$^{-1}$). The DLI--SHF curve is dominated by the CMB--SHF data which have the lowest uncertainty. The weighted mean tension between these two aggregate samples is $6.5\sigma$ (or $3.9\sigma$ after correlation adjustment). The vertical dashed line indicates the Planck 2018 value.}
    \label{fig:Inverse-Variance_Weighted_Sum_of_Gaussians}
\end{figure}

\subsubsection{Category 3: Pure Local \texorpdfstring{($H_0 = 71.03 \pm 0.69~\mathrm{km~s^{-1}~Mpc^{-1}}$)}{(H0 = 71.03 +/- 0.69 km s\textasciicircum-1 Mpc\textasciicircum-1)}}

The 16 cosmological-model-independent measurements (Megamasers, Cosmic Chronometers, EPM) yield an intermediate value of $H_0 = 71.03 \pm 0.69$~km~s$^{-1}$~Mpc$^{-1}$. Crucially, the removal of the $\Lambda$CDM shape assumption results in a central value shift of $\Delta H_0 = +3.42$~km~s$^{-1}$~Mpc$^{-1}$ relative to Category~2. This shift is visible in Figure~\ref{fig:Gaussian_probability_distributions} as a broader distribution positioned between the $\Lambda$CDM-dependent results and the Distance Ladder.

The reduced chi-squared $\chi^2_\nu = 0.77$ is lower than unity, indicating either conservative error estimates or the presence of correlations. This category exhibits significant internal heterogeneity: the Vogl et al.~\cite{Vogl2024} tailored EPM measurement ($H_0 = 74.9 \pm 2.7$)  strongly favors  the Distance Ladder value, while most Cosmic Chronometer measurements favor values closer to Planck. This heterogeneity suggests that Category~3 may itself contain distinct subpopulations, a possibility we explore further in Section~\ref{sec:model_dependence}.

\subsubsection{Category 4: CMB Sound Horizon Free (CMB--SHF)\texorpdfstring{($H_0 = 69.07 \pm 0.44~km~s^{-1}~Mpc^{-1}$)}{(H0 = 69.07 +/- 0.44 km s\textasciicircum-1 Mpc\textasciicircum-1)}} 

The 9 measurements utilizing CMB data without the Sound Horizon yield $H_0 = 69.07 \pm 0.44$~km~s$^{-1}$~Mpc$^{-1}$. This confirms that early-universe constraints on matter density (via the equality scale $k_{\rm eq}$) favor a low $H_0$ regardless of the $r_s$ calibration~\cite{Pogosian2024}.

The reduced chi-squared $\chi^2_\nu = 0.89$ is very close to unity, indicating good internal consistency among the $N = 9$ measurements in this category. However, given that several measurements share common methodologies and partially overlapping datasets---particularly the Pogosian et al.~\cite{Pogosian2024,Pogosian2025priv} and Garc\'ia Escudero et al.~\cite{GarciaEscudero:2025} results, which all employ Sound-Horizon-Free BAO analyses with DESI data---we examine potential correlations in Appendix~\ref{app:correlations}. In Figure~\ref{fig:Gaussian_probability_distributions}, this distribution overlaps significantly with CMB--SHF measurements, reinforcing the consistency between early-universe physics and local data when $\Lambda$CDM is assumed.

\subsection{Comparison: Distance Ladder vs.\ Combined Distance Ladder Independent/Sound Horizon Free (DLI--SHF)}
\label{subsec:ladder_vs_onestep}

To test the robustness of the ``Early vs.\ Late'' narrative, we combine all 58 DLI--SHF from Categories~2, 3, and 4 into a single ensemble. This combined sample represents the entirety of the evidence against the Distance Ladder that does not rely on the $r_s$ standard ruler.

The weighted mean of this combined sample is:
\begin{equation}
H_0^{\rm All\text{-}Ladder\text{-}Indep} = 69.37 \pm 0.34\,\mathrm{km\,s^{-1}\,Mpc^{-1}}.
\end{equation}
This distribution is shown as the black curve in Figure~\ref{fig:Gaussian_probability_distributions} and the red curve in Figure~\ref{fig:Inverse-Variance_Weighted_Sum_of_Gaussians}, contrasting sharply with the Distance Ladder (blue curve). Figure~\ref{fig:Inverse-Variance_Weighted_Sum_of_Gaussians} displays the probability density functions (PDFs) for the two measurement categories. Each individual measurement is represented as a Gaussian distribution, and the combined PDF is constructed as an inverse-variance weighted sum:
\begin{equation}
p_{\mathrm{raw}}(H_0) = \sum_{i=1}^{N} w_i \cdot \mathcal{N}(H_0 \mid H_{0,i}, \sigma_i)
\label{eq:pdf_raw}
\end{equation}
where $\mathcal{N}(x \mid \mu, \sigma)$ is the normal distribution:
\begin{equation}
\mathcal{N}(x \mid \mu, \sigma) = \frac{1}{\sigma\sqrt{2\pi}} \exp\left( -\frac{(x-\mu)^2}{2\sigma^2} \right)
\label{eq:gaussian}
\end{equation}
The final normalized PDF is obtained by dividing by the total area:
\begin{equation}
p(H_0) = \frac{p_{\mathrm{raw}}(H_0)}{\displaystyle\int_{-\infty}^{\infty} p_{\mathrm{raw}}(H_0') \, dH_0'}
\label{eq:pdf_norm}
\end{equation}
The two curves exhibit significant overlap despite the tension between their weighted means. This occurs because the image reflects the spread of individual measurement uncertainties $\sigma_i$ (typically 2--10 $\mathrm{km\,s^{-1}\,Mpc^{-1}}$), whereas the uncertainty on the weighted mean is much smaller. Consequently, the PDF width is governed by the individual $\sigma_i$ values, while the tension is calculated using the combined $\sigma_{\hat{H}_0}$ of Eq.~\eqref{eq:weighted_mean}.

This distinction explains why the curves overlap visually even when the weighted means differ by several standard deviations. The tension between the Distance Ladder and the Combined DLI--SHF sample is:
\begin{equation}
\Delta H_0 = 72.73 - 69.37 = 3.36\,\mathrm{km\,s^{-1}\,Mpc^{-1}}.
\end{equation}
Combining the uncertainties in quadrature ($\sigma_{\rm diff} = \sqrt{0.39^2 + 0.34^2} \approx 0.517$~km~s$^{-1}$~Mpc$^{-1}$), the significance of the discrepancy is:
\begin{equation}
n_\sigma = \frac{3.36}{0.517} \approx 6.5\sigma.
\end{equation}

This $6.5\sigma$ tension exceeds the standard Planck vs.\ SH0ES discrepancy ($\sim 5\sigma$). It suggests that the Distance Ladder is in conflict not just with the CMB Sound Horizon, but with the aggregate of essentially all other cosmological probes that are independent of $r_s$.

\subsubsection{Comparison with Previous Results}

Table~\ref{tab:results_summary} includes a comparison with our 2024 analysis~\cite{Perivolaropoulos2024}. Despite the significant expansion of the sample (from 53 to 88 measurements) and the refinement of the classification scheme (from 2 to 4 categories), the core results remain consistent:
\begin{itemize}
    \item The Distance Ladder weighted mean has shifted slightly from $72.8 \pm 0.5$ to $72.73 \pm 0.39$~km~s$^{-1}$~Mpc$^{-1}$, with improved precision due to the inclusion of JWST measurements.
    \item The DLI--SHF weighted mean has shifted from $68.3 \pm 0.5$ to $69.37 \pm 0.34$~km~s$^{-1}$~Mpc$^{-1}$, remaining consistent within uncertainties.
    \item The tension has increased from $\sim 5\sigma$ to $6.5\sigma$, primarily due to reduced uncertainties rather than a change in central values.
\end{itemize}

The stability of these results across independent analyses with different sample compositions strengthens confidence in the robustness of the ``Distance Ladder vs.\ The Rest'' characterization of the Hubble tension.

\subsubsection{Impact of Correlations}

The $6.5\sigma$ tension quoted above assumes statistical independence between measurements. As detailed in Appendix~\ref{app:correlations}, accounting for realistic correlations within each category increases the combined uncertainties. Using correlation coefficients estimated from shared calibration data and common methodology ($\bar{\rho} \approx 0.3$--$0.5$ for Distance Ladder; $\bar{\rho} \approx 0.1$--$0.2$ for DLI--SHF), we obtain:
\begin{align}
H_0^{\rm Ladder, corr} &= 72.30 \pm 0.57\,\mathrm{km\,s^{-1}\,Mpc^{-1}}, \\
H_0^{\rm Ladder\text{-}Indep, corr} &= 69.52 \pm 0.42\,\mathrm{km\,s^{-1}\,Mpc^{-1}}.
\end{align}
The correlation-adjusted tension is:
\begin{equation}
n_\sigma^{\rm corr} = \frac{72.30 - 69.52}{\sqrt{0.57^2 + 0.42^2}} \approx 3.9\sigma.
\end{equation}
While reduced from $6.5\sigma$, this tension remains highly significant and robust across the range of plausible correlation assumptions ($3.9$--$4.3\sigma$ for reasonable correlation models).

\subsection{Internal Tension: Model Dependence}
\label{subsec:internal_tension}

Beyond the external tension with the Distance Ladder, we identify a notable internal tension among the Distance Ladder Independent/Sound Horizon Free(DLI--SHF)  measurements based on model assumptions. Comparing Category~2 (Local $\Lambda$CDM) with Category~3 (Pure Local), we find:
\begin{equation}
\Delta H_0^{\rm internal} = 71.03 - 67.61 = 3.42\,\mathrm{km\,s^{-1}\,Mpc^{-1}}.
\end{equation}
The statistical significance is:
\begin{equation}
n_\sigma^{\rm internal} = \frac{3.42}{\sqrt{0.69^2 + 0.96^2}} \approx 2.9\sigma.
\label{eq:internal_tension}
\end{equation}

This $2.9\sigma$ internal tension, while not definitive, is suggestive of a systematic effect: imposing the $\Lambda$CDM model on local data appears to pull the inferred $H_0$ downward by $\sim 3.4$~km~s$^{-1}$~Mpc$^{-1}$. This shift accounts for more than half of the total Hubble tension ($\sim 6$~km~s$^{-1}$~Mpc$^{-1}$ between Distance Ladder and Planck).

We caution that this internal tension should be interpreted carefully for several reasons:
\begin{enumerate}
    \item \textbf{Look-elsewhere effect:} We are comparing multiple categories, which increases the probability of finding a $\sim 2.9\sigma$ discrepancy by chance.
    \item \textbf{Category heterogeneity:} As noted above, Category~3 contains measurements spanning a wide range of $H_0$ values (from $\sim 66$ to $\sim 75$~km~s$^{-1}$~Mpc$^{-1}$).
    \item \textbf{Sample size:} Category~3 contains only 16 measurements, making the weighted mean sensitive to individual high-precision entries.
\end{enumerate}

Nevertheless, the model-dependent shift is clearly observable in the offset between the Category~2 and Category~3 distributions in Figure~\ref{fig:Gaussian_probability_distributions}, and its physical implications are explored further in Section~\ref{sec:model_dependence}.

\subsection{Summary of Statistical Results}
\label{subsec:results_summary}

The key statistical findings of this section are:
\begin{enumerate}
    \item The Distance Ladder yields $H_0 = 72.73 \pm 0.39$~km~s$^{-1}$~Mpc$^{-1}$, consistent with SH0ES and JWST measurements.
    \item The combined DLI--SHF measurements yield $H_0 = 69.37 \pm 0.34$~km~s$^{-1}$~Mpc$^{-1}$, slightly elevated compared to our previous results but closer to Planck.
    \item The tension between these two ensembles is $6.5\sigma$ (or $3.9\sigma$ after correlation adjustment).
    \item Local measurements assuming $\Lambda$CDM (Category~2) yield systematically lower $H_0$ than cosmological-model-independent local measurements (Category~3), with a $2.9\sigma$ internal tension.
    \item These results are consistent with, and strengthen, the conclusions of our 2024 analysis.
\end{enumerate}

The implications of these findings---particularly the role of model assumptions and the apparent bimodality in the $H_0$ distribution---are explored in the following section.
%======================================================================
%\section{The Impact of Model Assumptions}
%\label{sec:model_dependence}
%======================================================================
% Goals for this section:
% 1. Compare Category 2 (LCDM) vs Category 3 (Model Independent).
% 2. Discuss the finding that assuming LCDM shape pulls H0 lower.
% 3. Discuss the "Bimodality" observed in the one-step data (from the previous draft) in the context of model dependence.

%======================================================================
\section{The Impact of Model Assumptions}
\label{sec:model_dependence}
%======================================================================

The results of Section~\ref{sec:results} reveal a striking feature of the Hubble tension landscape: the discrepancy is not merely between the Distance Ladder and the Early Universe, but also manifests as an internal tension within the domain of Distance Ladder Independent/Sound Horizon Free (DLI--SHF) measurements. Specifically, the statistical compatibility of a measurement with the Planck value depends critically on whether the standard $\Lambda$CDM expansion history is assumed during the inference process. In this section, we dissect this model dependence, examine the critical role of specific high-precision measurements, and explore the physical implications of the observed bimodality in the $H_0$ distribution.

\subsection{Implicit Bias of the \texorpdfstring{$\Lambda$}{Λ}CDM Expansion History}
\label{subsec:lcdm_bias}

The most significant finding of our four-category classification is the $2.9\sigma$ tension between local DLI--SHF measurements that assume a $\Lambda$CDM cosmology (Category~2: $H_0 = 67.61 \pm 0.96$~km~s$^{-1}$~Mpc$^{-1}$) and those that do not (Category~3: $H_0 = 71.03 \pm 0.69$~km~s$^{-1}$~Mpc$^{-1}$).

Measurements in Category~2, such as Gravitational Wave standard sirens~\cite{Abbott2021} and standard analyses of Strong Lensing time delays~\cite{Shajib2023}, typically span the redshift range $0 < z < 1.5$. To extract $H_0$ from the observed luminosity distances or time delays, these methods must integrate the Hubble parameter $H(z)$:
\begin{equation}
d_L(z) = (1+z) \int_0^z \frac{c\, dz'}{H(z')}.
\label{eq:luminosity_distance}
\end{equation}
By fixing the functional form of $H(z)$ to that of flat $\Lambda$CDM (Eq.~\ref{eq:lcdm_hz}), often with $\Omega_m$ fixed to $\approx 0.3$ or marginalized over a narrow prior, these analyses effectively impose a rigid shape constraint on the expansion history. The inferred $H_0$ is then the normalization that best fits the data given this assumed shape.

Our analysis shows that this $\Lambda$CDM shape constraint systematically prefers lower $H_0$ values ($H_0 \approx 67$~km~s$^{-1}$~Mpc$^{-1}$), aligning almost perfectly with the Planck CMB inference~\cite{Planck2018}. This demonstrates a remarkable self-consistency of the $\Lambda$CDM model: if one assumes the $\Lambda$CDM background evolution holds true at late times, local data (excluding the Distance Ladder) yields results consistent with early-universe physics, even without using the Sound Horizon~\cite{Banik:2024vbz}.

However, when this shape constraint is released (Category~3), the inferred $H_0$ rises by $\Delta H_0 \approx +3.4$~km~s$^{-1}$~Mpc$^{-1}$ to $\approx 71$~km~s$^{-1}$~Mpc$^{-1}$. This shift has two possible interpretations:

\begin{enumerate}
    \item \textbf{$\Lambda$CDM is correct, and Category~3 has larger systematics:} Cosmological-model-independent methods may suffer from unrecognized systematic errors that bias $H_0$ high. The lack of a shape constraint allows more freedom for systematics to propagate into the final result.
    
    \item \textbf{The true $H(z)$ deviates from $\Lambda$CDM:} If the actual expansion history differs from the $\Lambda$CDM prediction at $z < 1$ (e.g., due to a late-time gravitational transition, dynamical dark energy, or modified gravity~\cite{Perivolaropoulos2024,Hogas:2025ahb,Plaza:2025gcv,Odintsov:2025jfq,Pan:2025psn,Akarsu:2025gwi}), forcing a $\Lambda$CDM fit would bias the inferred $H_0$ downward to compensate for the shape mismatch.
\end{enumerate}
A related possibility is that the tension reflects not a global modification of $H(z)$, but rather a localized change in the physics governing distance calibrators. Refs~\cite{Desmond:2019ygn,Desmond:2020wep,Marra2021,Ruchika2025,Perivolaropoulos:2025gzo} investigated whether a local variation in fundamental physics---such as a change in the fine-structure constant or gravitational strength within the calibration volume---could reconcile the SH0ES measurement with other probes without requiring modifications to the global expansion history.The current data do not definitively distinguish between these scenarios. A detailed comparison of late-transition versus smooth H(z) deformation models has been presented in~\cite{Alestas:2021luu}. Earlier reanalyses of SH0ES data exploring additional degrees of freedom also found evidence for such model-dependent effects~\cite{Perivolaropoulos:2022khd}.

\subsection{The Vogl et al. Tailored EPM: A Critical Test Case}
\label{subsec:vogl_epm}

Among the Category~3 measurements, the Vogl et al.~\cite{Vogl2024} tailored Expanding Photosphere Method (EPM) result deserves particular attention. This measurement yields $H_0 = 74.9 \pm 2.7$~km~s$^{-1}$~Mpc$^{-1}$  (with systematic uncertainties — by the authors' own estimate comparable to the statistical errors — taken into account) is fully consistent with the Distance Ladder and in $\sim 2.9\sigma$ tension with the Planck value---yet it is completely independent of both the Distance Ladder calibration chain and cosmological model assumptions.

The EPM determines distances to Type~II supernovae geometrically by combining measurements of the photospheric angular size (from flux and temperature) with the photospheric velocity (from spectral line Doppler shifts). The ``tailored'' approach of Vogl et al. advances this technique by using detailed non-local thermodynamic equilibrium (NLTE) radiative transfer models customized to individual supernovae, rather than the empirical dilution factors traditionally employed. This reduces model-dependent systematics and achieves significantly smaller distance uncertainties.

The significance of this measurement for the Hubble tension debate is threefold:

\begin{enumerate}
    \item \textbf{Independence from SNe~Ia physics:} Unlike the Distance Ladder, which relies on the standardizability of SNe~Ia, the tailored EPM uses SNe~II---an entirely different supernova class with distinct explosion physics.
    
    \item \textbf{Independence from calibrator physics:} The EPM does not require Cepheids, TRGB, or any other intermediate calibrator. It provides a direct geometric distance to the supernova.
    
    \item \textbf{Independence from $\Lambda$CDM:} The method measures distances without integrating $H(z)$, making it insensitive to the assumed cosmological model.
\end{enumerate}

If the Distance Ladder's high $H_0$ were solely due to systematic errors in Cepheid photometry, metallicity corrections, or SNe~Ia standardization, one would not expect an independent method like the tailored EPM to yield a consistent result. The agreement between Vogl et al. and the Distance Ladder suggests that either: (a) both methods share a common, as-yet-unidentified systematic, or (b) the high $H_0$ value reflects the true local expansion rate.

However, we note important caveats. The tailored EPM relies on assumptions about supernova explosion physics, ejecta composition, and atmospheric structure. The current sample is limited to a relatively small number of well-observed SNe~II. Further validation with larger samples and independent atmosphere codes will be essential to confirm this result. 

\subsection{The Megamaser Perspective}
\label{subsec:megamasers}

A similar pattern emerges from Megamaser measurements. The Barua et al.~\cite{Barua2025} frequentist analysis of the Megamaser Cosmology Project (MCP) sample yields $H_0 = 73.5 \pm 3.0$~km~s$^{-1}$~Mpc$^{-1}$, again consistent with the Distance Ladder.

Megamaser distances are determined geometrically from the Keplerian dynamics of circumnuclear water maser disks, providing angular diameter distances that are in principle completely independent of stellar physics. However, Megamaser $H_0$ measurements are sensitive to the treatment of peculiar velocities, which become significant at the low redshifts ($z \lesssim 0.05$) where most masers hosts reside.

In our classification, we assigned the Barua et al~\cite{Barua2025} measurement to Category 2 because it relies primarily on Keplerian disk modeling without adopting $\Lambda$CDM-based peculiar velocity priors but relies on the $\Lambda$CDM prediction for the angular diameter distance. Indeed, when $\Omega_m$ is left as a free parameter rather than fixed to the Planck value, the best-fit $H_0$ drops to $\sim 63$~km~s$^{-1}$~Mpc$^{-1}$ with uncertainties approximately five times larger, highlighting the model dependence of the headline result. In contrast, earlier analyses that adopted peculiar velocity corrections from the SH0ES flow model from Pesce et al.~\cite{Pesce2020} would be more appropriately classified as having some distance ladder dependence. 

The earlier Megamaser measurements by Kuo et al.~\cite{Kuo2013} ($H_0 = 68 \pm 9$~km~s$^{-1}$~Mpc$^{-1}$) and Gao et al.~\cite{Gao2015} ($H_0 = 66 \pm 6$~km~s$^{-1}$~Mpc$^{-1}$) yielded lower values, though with larger uncertainties. The evolution toward higher $H_0$ in recent Megamaser analyses may reflect improved disk modeling, better peculiar velocity treatments, or statistical fluctuations in small samples.

\subsection{Bimodality in the Distance Ladder Independent/Sound Horizon Free (DLI--SHF) Distribution}
\label{subsec:bimodality}

Although Figure~\ref{fig:Inverse-Variance_Weighted_Sum_of_Gaussians} does not exhibit an obvious bimodal structure---the distribution of DLI--SHF measurements appears reasonably concentrated around a single peak---a more detailed examination reveals a subtle but significant effect.

To further explore the underlying structure of the DLI--SHF $H_0$ measurements, demonstrated also in figure~\ref{fig:Gaussian_probability_distributions},  we employ a weighted Kernel Density Estimation (KDE) approach. Unlike a simple inverse-variance weighted sum of Gaussians, where each measurement contributes a Gaussian with width equal to its uncertainty $\sigma_i$, the KDE method assigns a fixed bandwidth $h$ to all measurements while weighting their contributions according to their precision. The weighted KDE is defined as:
\begin{equation}
    \hat{f}(x) = \sum_{i=1}^{n} w_i \cdot \frac{1}{\sqrt{2\pi}\,h} \exp\left(-\frac{(x - x_i)^2}{2h^2}\right),
    \label{eq:kde}
\end{equation}
where $x_i$ is the measured $H_0$ value, $h$ is the bandwidth (kernel width), and $w_i$ are the normalized inverse-variance weights:
\begin{equation}
    w_i = \frac{1/\sigma_i^2}{\sum_{j=1}^{n} 1/\sigma_j^2}.
    \label{eq:weights}
\end{equation}

This approach ensures that high-precision measurements contribute proportionally through their weights, but do not dominate the distribution through both weight \textit{and} narrow kernel width simultaneously---a problem that arises with the standard inverse-variance weighted sum of Gaussians.

For this analysis, we exclude Category~4 measurements (CMB Sound Horizon Free), as these rely on CMB-derived data that provide exceptionally high precision. The three most precise Category~4 measurements---Pogosian et al.~\cite{Pogosian2024,Pogosian2025priv} (DESI: $69.37 \pm 0.65$~km~s$^{-1}$~Mpc$^{-1}$; pre-DESI: $68.05 \pm 0.94$~km~s$^{-1}$~Mpc$^{-1}$) and Garcia et al.~\cite{GarciaEscudero:2025} ($70.03 \pm 0.97$~km~s$^{-1}$~Mpc$^{-1}$)---would alone account for approximately $65\%$ of the total inverse-variance weight, clustering near $H_0 \approx 69$~km~s$^{-1}$~Mpc$^{-1}$ and obscuring any underlying structure in the remaining measurements.

The bandwidth $h$ is a critical parameter controlling the smoothness of the KDE. Silverman's rule of thumb~\cite{Silverman1986}, which minimizes the mean integrated squared error assuming a unimodal Gaussian distribution, yields $h \approx 2.2$~km~s$^{-1}$~Mpc$^{-1}$ for our data. However, this criterion is known to oversmooth multimodal distributions. Given the \textit{a priori} distinction between Category~2 ($\Lambda$CDM-dependent) and Category~3 (model-independent) measurements, a smaller bandwidth is justified to reveal potential substructure.

We find that bimodal structure is clearly visible for bandwidths in the range $h \in [1.1, 1.8]$~km~s$^{-1}$~Mpc$^{-1}$. For $h < 1.1$, additional peaks emerge due to over-resolution of statistical fluctuations; for $h > 1.8$, the two peaks merge into a single mode with a shoulder. We adopt $h = 1.5$~km~s$^{-1}$~Mpc$^{-1}$, an intermediate value within the robust range, to balance resolution and smoothing.

\begin{figure}[htbp]
    \centering
    \includegraphics[width=\columnwidth]{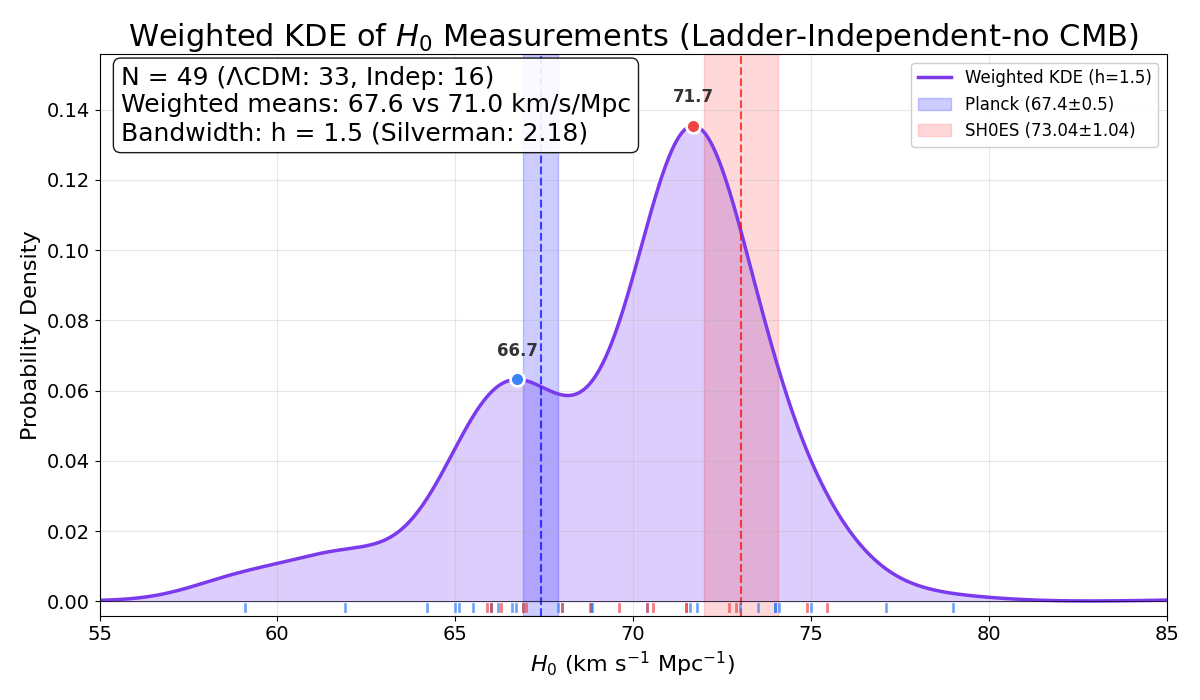}
    \caption{Weighted Kernel Density Estimation (KDE) of DLI--SHF $H_0$ measurements, 
    revealing the bimodal structure between Category 2 ($\Lambda$CDM-dependent) and 
    Category 3 (model-independent) measurements. We adopt a bandwidth $h = 1.5$ 
    km\,s$^{-1}$\,Mpc$^{-1}$, below the Silverman optimal value ($h \approx 2.2$), 
    to prevent the distribution from being dominated by a few high-precision 
    measurements. The bimodal structure remains robust across the range 
    $h \in [1.1, 1.8]$; we select an intermediate value to balance resolution 
    and smoothing. Weights are assigned as $w_i = 1/\sigma_i^2$, while the 
    bandwidth is held constant across all measurements, ensuring that precise 
    measurements contribute proportionally without overwhelming the distribution 
    through both weight and kernel width.}
    \label{fig:Bimodality_of_one_step}
\end{figure}

The resulting distribution, shown in Figure~\ref{fig:Bimodality_of_one_step}, exhibits two distinct peaks corresponding to the weighted means of Category~2 ($\bar{H}_0 = 67.6$~km~s$^{-1}$~Mpc$^{-1}$) and Category~3 ($\bar{H}_0 = 71.0$~km~s$^{-1}$~Mpc$^{-1}$) measurements, with a separation of $\Delta H_0 = 3.4$~km~s$^{-1}$~Mpc$^{-1}$. This bimodal structure suggests a systematic difference between $\Lambda$CDM-dependent and cosmological-model-independent not Distance Ladder methods that warrants further investigation.

\subsection{Physical Interpretation of the Bimodality}
\label{subsec:bimodality_interpretation}

If the bimodality is real and not an artifact of small-number statistics, it suggests that the ``Hubble Tension'' is not characterized by a single discrepant value but rather by a bifurcation of results based on methodology. The two peaks can be characterized as follows:

\begin{enumerate}
    \item \textbf{The Low-$H_0$ Peak ($\sim 67$~km~s$^{-1}$~Mpc$^{-1}$):} Populated predominantly by:
    \begin{itemize}
        \item Category~2 measurements that assume $\Lambda$CDM (FRBs, GW sirens, most Strong Lensing)
        \item Cosmic Chronometer measurements in Category~3
    \end{itemize}
    These methods either explicitly assume $\Lambda$CDM or rely on physics that is ultimately calibrated by the same underlying cosmological model.
    
    \item \textbf{The High-$H_0$ Peak ($\sim 73$~km~s$^{-1}$~Mpc$^{-1}$):} Populated by:
    \begin{itemize}
        \item Specific measurements: Megamasers~\cite{Barua2025} and tailored EPM~\cite{Vogl2024}
    \end{itemize}
    These methods are characterized by their reliance on local geometric or kinematic physics.
\end{enumerate}

This bifurcation pattern suggests a reformulation of the Hubble tension: the discrepancy may fundamentally be between \emph{model-dependent} methods (which yield low $H_0$) and \emph{local astrophysics dependent or cosmological-model-independent} methods (which yield high $H_0$), rather than between early and late times.

\subsection{The Role of High-Precision Anchors}
\label{subsec:anchors}

The tension landscape is significantly shaped by a small number of high-precision measurements that serve as ``anchors'' for the two solutions:

\textbf{High-$H_0$ anchors:}
\begin{itemize}
    \item Riess et al.~\cite{Riess2025} JWST Cepheids: $H_0 = 73.49 \pm 0.93$~km~s$^{-1}$~Mpc$^{-1}$
    \item Zhang et al.~\cite{Zhang2024} pairwise velocities: $H_0 = 75.5 \pm 3.8$~km~s$^{-1}$~Mpc$^{-1}$
    \item Vogl et al.~\cite{Vogl2024} tailored EPM: $H_0 = 74.9 \pm 2.7$~km~s$^{-1}$~Mpc$^{-1}$
\end{itemize}

\textbf{Low-$H_0$ anchors:}
\begin{itemize}
    \item Pogosian et al.~\cite{Pogosian2025priv} DESI BAO (no $r_s$): $H_0 = 69.37 \pm 0.65$~km~s$^{-1}$~Mpc$^{-1}$
    \item Kelly et al.~\cite{Kelly2023} SN Refsdal: $H_0 = 66.6 \pm 3.7$~km~s$^{-1}$~Mpc$^{-1}$
\end{itemize}
Multi-parameter transition analyses of SH0ES data have explored how calibrator parameters may vary across the distance ladder, potentially contributing to the observed bifurcation. Evidence for possible transitions in Cepheid-SNe Ia calibrator parameters has been previously identified in the SH0ES sample~\cite{Perivolaropoulos:2021bds,Sharma:2025ucg}. Recent work has also highlighted the subtle statistical effects in Distance Ladder analyses, including distance priors and selection effects, that could impact such comparisons \cite{Desmond:2025ggt}.
The weighted means of each category are heavily influenced by these anchor measurements. Consequently, resolving the Hubble tension requires either: (a) identifying systematic errors in one or more anchor measurements, or (b) explaining why different methodologies yield systematically different results.

\subsection{Category 3 as the Critical Adjudication Ground}
\label{subsec:category3_adjudication}

The ``Pure Local category (Category~3) serves as the critical test for distinguishing between the two interpretations of the Hubble tension. Currently, this category yields an intermediate value ($H_0 = 71.03 \pm 0.69$~km~s$^{-1}$~Mpc$^{-1}$), but it is internally heterogeneous:

\begin{itemize}
    \item \textbf{High-$H_0$ subset:} Vogl EPM ($74.9 \pm 2.7$), Palmese GW170817 (75.46±5.4)
    \item \textbf{Low-$H_0$ subset:} Zhang CC+HII ($65.9 \pm 3.0$), Li lensing ($66.3 \pm 3.7$)
\end{itemize}

The future evolution of Category~3 will be decisive:

\begin{enumerate}
    \item \textbf{If Category~3 converges toward Category~1 (Distance Ladder):} This would strongly imply that the $\Lambda$CDM model assumption used in Category~2 is the source of the bias. Such convergence would suggest that the true late-time expansion history deviates from $\Lambda$CDM, requiring new physics (e.g., dynamical dark energy, modified gravity, or a late-time gravitational transition)such as the $\Lambda_s$CDM model~\cite{akarsu2024,Akarsu2025} 
    \item \textbf{If Category~3 converges toward Category~2 ($\Lambda$CDM):} This would suggest that the Distance Ladder contains a unique, unrecognized systematic error affecting all its calibrators (Cepheids, TRGB, JAGB, Miras) in a similar way. Potential culprits include metallicity effects, dust extinction corrections, or environmental dependencies of SNe~Ia.
\end{enumerate}

Currently, the intermediate position of Category~3, combined with its internal heterogeneity, does not clearly favor either scenario. However, the existence of multiple independent cosmological-model-independent measurements (EPM, Megamasers) that support the high-$H_0$ solution is suggestive.

\subsection{Summary: The Model-Dependence Effect}
\label{subsec:model_summary}

The key findings of this section can be summarized as follows:

\begin{enumerate}
    \item Imposing $\Lambda$CDM assumptions on local kinematic data reduces the inferred $H_0$ by approximately $3.4$~km~s$^{-1}$~Mpc$^{-1}$ (from Category~3's $71.03$ to Category~2's $67.61$).
    
    \item This shift accounts for more than half of the total Hubble tension ($\sim 6$~km~s$^{-1}$~Mpc$^{-1}$ between Distance Ladder and Planck).
    
    \item The distribution shows evidence for bimodality %when high-precision anchoring measurements are removed%,
    with peaks corresponding to model-dependent ($\sim 67$) and cosmological-model-independent ($\sim 73$) methods.
    
    \item Specific cosmological-model-independent measurements (Vogl EPM) yield high $H_0$ values consistent with the Distance Ladder, suggesting the high value may reflect true local physics rather than distance ladder systematics.
    
    \item The future convergence or divergence of Category~3 measurements will be critical for determining whether the tension originates from $\Lambda$CDM model bias or distance ladder systematics.
\end{enumerate}

These findings support a recharacterization of the Hubble tension as primarily a ``Distance Ladder vs.\ The Rest'' crisis, with a secondary ``Model-Dependent vs.\ Cosmological-model-Independent'' tension that may hold the key to its ultimate resolution.

%======================================================================
%\section{Discussion and Conclusions}
%\label{sec:conclusions}
%======================================================================
% Goals for this section:
% 1. Synthesize the main conclusion: Tension is "Distance Ladder vs The Rest".
% 2. Highlight that "Local" does not equal "High H0" (as seen in Cat 2 & 4).
% 3. Summarize the internal tension regarding H(z) shape assumptions.
% 4. Final remarks on systematics in the Distance Ladder vs New Physics.

%======================================================================
\section{Discussion and Conclusions}
\label{sec:conclusions}
%======================================================================

The ``Hubble Tension'' has long been framed as a cosmological crisis pitting the Early Universe (calibrated by the Sound Horizon) against the Late Universe (measured by the distance ladder). In this work, we have dissected this narrative by analyzing 88 measurements of $H_0$ that are strictly independent of the CMB Sound Horizon scale. Our findings suggest that the crisis is not necessarily a temporal one, but rather a methodological one. In this concluding section, we synthesize our results, discuss their implications for proposed solutions to the tension, and outline the observations needed to achieve resolution.

\subsection{The ``Distance Ladder vs.\ The Rest'' Crisis}
\label{subsec:ladder_crisis}

Our primary conclusion is that the Hubble tension is most accurately characterized as a discrepancy between the Distance Ladder methodology and essentially all other cosmological probes, regardless of whether those probes utilize early- or late-time data.

As detailed in Section~\ref{sec:results}, the Distance Ladder (Category~1) yields a robust high value of $H_0 = 72.73 \pm 0.39$~km~s$^{-1}$~Mpc$^{-1}$. In stark contrast, when we combine all 58 independent measurements from Categories~2, 3, and 4---comprising Gravitational Waves, Strong Lensing, Cosmic Chronometers, Megamasers, FRBs, and Sound-Horizon-Free CMB data---we obtain a combined value of $H_0 = 69.37 \pm 0.34$~km~s$^{-1}$~Mpc$^{-1}$.

The resulting $6.5\sigma$ tension ($3.9\sigma$ after correlation adjustment) between the Distance Ladder and the Distance Ladder Independent/Sound Horizon Free (DLI--SHF) ensemble exceeds the standard Planck vs.\ SH0ES discrepancy ($\sim 5\sigma$). This indicates that the high value of $H_0$ is a unique feature of the three-rung calibration chain (involving Cepheids/TRGB/Miras $\to$ SNe~Ia) and is not supported by the broader consensus of Sound-Horizon-Free cosmology.

This conclusion is robust across our analyses:
\begin{itemize}
    \item It persists when correlations within categories are accounted for (Section~\ref{subsec:ladder_vs_onestep}).
    \item It is consistent with our 2024 analysis~\cite{Perivolaropoulos2024}, despite the significant expansion of the sample.
    \item It holds regardless of which specific measurements are included or excluded (with the exception of complete removal of entire methodological classes).
\end{itemize}

% Add this to your conclusions section
% Note: Requires \usepackage{xcolor}, \usepackage{booktabs}, \usepackage{graphicx} in preamble

\subsection{Robustness of Results}

To assess the sensitivity of our results to methodological choices in measurement categorization, we performed a comprehensive robustness analysis\footnote{We thank Dan Scolnic for suggesting the examination of the categorization sensitivity for these specific cases to test the robustness of the internal tension.}. We considered a more conservative approach to categorization by: (i) adopting alternative reported values for measurements with multiple H$_0$ estimates (Liu et al.: 59.1 $\rightarrow$ 67.5 km s$^{-1}$ Mpc$^{-1}$; Dominguez et al.: 61.9 $\rightarrow$ 65.6 km s$^{-1}$ Mpc$^{-1}$), (ii) removing potentially redundant measurements from the same observational data (Kelly/Li Refsdal analyses), and (iii) excluding measurements whose categorization could be debated (Zhang et al. 2022 cosmic chronometer analysis). Table~\ref{tab:robustness} summarizes the impact of these changes on our tension estimates.

\begin{table}[b]
\centering
\caption{Sensitivity analysis of tension estimates to categorization choices. The main tension between Distance Ladder and DLI--SHF methods remains highly significant ($>5\sigma$) under all scenarios tested.}
\label{tab:robustness}
\scriptsize
\begin{tabular}{lcccc}
\toprule
Scenario & Cat2 & Cat3 & Tension & Tension \\
 & $H_0 \pm \sigma$ & $H_0 \pm \sigma$ & (DL-LI) & (C2-C3) \\
\midrule
Baseline (this work) & $67.61 \pm 0.96$ & $71.03 \pm 0.69$ & $6.49\sigma$ & $2.89\sigma$ \\
Liu: 59.1$\to$67.5 & $68.27 \pm 0.99$ & $71.03 \pm 0.69$ & $6.22\sigma$ & $2.29\sigma$ \\
Dominguez: 61.9$\to$65.6 & $68.35 \pm 1.01$ & $71.03 \pm 0.69$ & $6.20\sigma$ & $2.19\sigma$ \\
Remove Kelly (Refsdal) & $67.68 \pm 0.99$ & $71.03 \pm 0.69$ & $6.35\sigma$ & $2.78\sigma$ \\
Remove Li (Refsdal) & $67.61 \pm 0.96$ & $71.19 \pm 0.70$ & $6.35\sigma$ & $3.01\sigma$ \\
Remove Zhang 2022 & $67.61 \pm 0.96$ & $71.32 \pm 0.71$ & $6.32\sigma$ & $3.11\sigma$ \\
Liu + Dominguez & $69.15 \pm 1.04$ & $71.03 \pm 0.69$ & $6.01\sigma$ & $1.51\sigma$ \\
All changes (keep Li) & $69.37 \pm 1.09$ & $71.32 \pm 0.71$ & $5.86\sigma$ & $1.50\sigma$ \\
All changes (keep Kelly) & $69.15 \pm 1.04$ & $71.51 \pm 0.72$ & $5.86\sigma$ & $1.87\sigma$ \\
\bottomrule
\end{tabular}
\end{table}

The analysis reveals that our main result---the significant tension between Distance Ladder and all others measurements---is robust to all tested variations in categorization methodology. Even under the most conservative scenario combining all alternative choices, the main tension remains at $\sim 5.9\sigma$, well above the $5\sigma$ threshold for statistical significance. No individual change affects the main tension by more than $0.29\sigma$.

In contrast, the internal tension between Category 2 ($\Lambda$CDM-dependent) and Category 3 (cosmological-model-independent) measurements is more sensitive to these methodological choices, varying from $1.50\sigma$ to $3.11\sigma$ depending on the specific categorization adopted.

\subsection{Implications for Proposed Solutions}
\label{subsec:implications}

Our findings have significant implications for the major classes of proposed solutions to the Hubble tension:

\subsubsection{Early Dark Energy and Sound Horizon Modifications}

Models that resolve the tension by reducing the Sound Horizon at recombination (e.g., Early Dark Energy~\cite{Knox2020,Mirpoorian2025,Specogna:2025guo}, extra relativistic species, early modified gravity) predict that Sound-Horizon-Free measurements should align with the Distance Ladder. Our analysis shows the opposite: Distance Ladder Independent/Sound Horizon Free (DLI--SHF) (Categories~2, 3, and 4) yields a value closer to Planck, not to SH0ES.

This presents a significant challenge to early-universe solutions. If EDE or similar models were correct, we would expect:
\begin{itemize}
    \item Megamaser distances to yield $H_0 \approx 73$~km~s$^{-1}$~Mpc$^{-1}$ \checkmark(partially observed).
    \item GW standard sirens to yield $H_0 \approx 73$~km~s$^{-1}$~Mpc$^{-1}$ \ding{55} (observed: $\approx 68$)
    \item Strong lensing to yield $H_0 \approx 73$~km~s$^{-1}$~Mpc$^{-1}$ \ding{55} (observed: $\approx 67$--$71$)
    \item Sound-Horizon-Free BAO to yield $H_0 \approx 73$~km~s$^{-1}$~Mpc$^{-1}$ \ding{55} (observed: $\approx 69$)
\end{itemize}

The failure of most Sound-Horizon-Free probes to support the high-$H_0$ solution disfavors pure early-universe modifications as the complete explanation.

\subsubsection{Late-Time Modifications}

Models that modify the expansion history at late times (e.g., dynamical dark energy, late-time gravitational transitions, interacting dark energy) predict that the inferred $H_0$ should depend on whether $\Lambda$CDM is assumed in the analysis. Our observation of the Category~2 vs.\ Category~3 offset ($\Delta H_0 \approx 3.4$~km~s$^{-1}$~Mpc$^{-1}$) is qualitatively consistent with this prediction.

However, late-time solutions face their own challenges:
\begin{itemize}
    \item They must explain why the shift is only $\sim 3$~km~s$^{-1}$~Mpc$^{-1}$, not the full $\sim 6$~km~s$^{-1}$~Mpc$^{-1}$ tension.
    \item They must be consistent with other late-time probes (SNe~Ia Hubble diagrams, BAO, growth of structure) that show no dramatic deviation from $\Lambda$CDM.
    \item Recent DESI results~\cite{DESI2024,Giare:2025pzu,Gao:2024ily} suggesting evolving dark energy ($w_0 > -1$, $w_a < 0$) provide evidence for deviations from $\Lambda$CDM at late times. However, due to the strong anti-correlation between $w_0$ and $H_0$~\cite{Colgain2022}, such models with $w_0 > -1$ actually predict lower $H_0$ values, potentially exacerbating rather than alleviating the tension with the Distance Ladder. Moreover, the statistical significance of the DESI signal remains modest ($\sim 2$--$3\sigma$) and its robustness across different parameterizations has been questioned~\cite{Colgain:2024mtg}. Theoretical frameworks for such behavior include models where the effective gravitational constant $G_{eff}$ undergoes a rapid transition at late times, which can arise naturally in scalar-tensor theories. Recent re-examinations of such modified local physics scenarios using SH0ES data have refined the constraints on these models.  A particularly intriguing perspective recasts the Hubble tension as an $M$ crisis. Theodoropoulos \& Perivolaropoulos demonstrated that if the intrinsic luminosity of SNe Ia in the Hubble flow differs systematically from that in calibrator galaxies, the inferred $H_0$ would be biased without requiring modifications to fundamental cosmology~\cite{Marra2021,Perivolaropoulos:2025gzo,Theodoropoulos2021}. 
\end{itemize}

\subsubsection{Distance Ladder Systematics}

The isolation of the Distance Ladder as a $6.5\sigma$ outlier against 58 independent measurements points toward unrecognized systematics within the calibration chain. Potential systematic effects include:

\begin{itemize}
    \item \textbf{Cepheid systematics:} Metallicity dependence of the period-luminosity relation, crowding and blending effects, or extinction corrections.
    \item \textbf{SNe~Ia systematics:} Environmental dependencies (host galaxy mass, local star formation), evolution with redshift, or standardization procedure biases~\cite{Perivolaropoulos:2023iqj,Lopez-Hernandez:2025lbj,Teixeira:2025czm}.
    \item \textbf{Calibrator physics:} Differences between calibrators observed in anchor galaxies vs.\ SNe~Ia hosts.
    \item \textbf{Sample selection:} Potential biases in which galaxies are selected for Cepheid observation.
\end{itemize}

However, the ``systematics'' interpretation faces its own challenges:
\begin{itemize}
    \item Multiple independent calibrators (Cepheids, TRGB, JAGB, Miras, SBF) yield consistent high-$H_0$ values.
    \item JWST observations have largely confirmed HST Cepheid photometry~\cite{Riess2025}.
    \item Cosmological-model-independent methods like the tailored EPM~\cite{Vogl2024} support the high-$H_0$ solution.
\end{itemize}

\subsubsection{Hybrid Scenarios}

Given the complexity of the evidence, a hybrid scenario involving \emph{both} partial distance ladder systematics \emph{and} deviations from $\Lambda$CDM dynamics may be required. For example:
\begin{itemize}
    \item A $\sim 2$\% systematic in the distance ladder calibration (reducing its $H_0$ by $\sim 1.5$~km~s$^{-1}$~Mpc$^{-1}$)
    \item Combined with a late-time modification to $H(z)$ (accounting for the remaining $\sim 1.5$--$3$~km~s$^{-1}$~Mpc$^{-1}$)
\end{itemize}
Such a scenario is constrained but not excluded by current data.

\subsection{Falsifiability and Future Tests}
\label{subsec:falsifiability}

A robust scientific framework requires clear predictions that can be tested by future observations. We outline the key tests that would distinguish between the competing interpretations:

\subsubsection{What Would Confirm Distance Ladder Systematics?}

The ``systematics'' hypothesis would be strongly supported if:
\begin{enumerate}
    \item Category~3 measurements converge toward $H_0 \approx 67$~km~s$^{-1}$~Mpc$^{-1}$ as precision improves.
    \item Future Megamaser measurements yield $H_0 < 70$~km~s$^{-1}$~Mpc$^{-1}$ with reduced uncertainties.
    \item The tailored EPM, applied to larger SNe~II samples, yields $H_0 < 70$~km~s$^{-1}$~Mpc$^{-1}$.
    \item A specific systematic is identified in the distance ladder (e.g., a metallicity correction error) that shifts its $H_0$ down by $\sim 5$~km~s$^{-1}$~Mpc$^{-1}$.
    \item Independent geometric distance methods (e.g., detached eclipsing binaries in external galaxies) yield distances inconsistent with Cepheids.
\end{enumerate}

\subsubsection{What Would Confirm New Physics?}

The ``new physics'' hypothesis would be strongly supported if:
\begin{enumerate}
    \item Category~3 measurements converge toward $H_0 \approx 73$~km~s$^{-1}$~Mpc$^{-1}$ as precision improves.
    \item Multiple independent cosmological-model-independent methods (EPM, Megamasers, future techniques) consistently yield $H_0 > 72$~km~s$^{-1}$~Mpc$^{-1}$.
    \item GW standard sirens, analyzed without assuming $\Lambda$CDM (e.g., using cosmographic methods), yield higher $H_0$ than $\Lambda$CDM-based analyses.
    \item Direct evidence for deviations from $\Lambda$CDM at $z < 1$ emerges from other probes (e.g., DESI dark energy constraints, growth rate measurements).
    \item A theoretical model is developed that simultaneously explains the Hubble tension, the Category~2 vs. Category~3 offset, and passes all other cosmological tests.
\end{enumerate}

\subsection{Observational Priorities}
\label{subsec:priorities}

Based on our analysis, we identify the following observational priorities for resolving the Hubble tension:

\begin{enumerate}
    \item \textbf{Expand Category~3 (Pure Local):} This category currently contains only 16 measurements with significant internal heterogeneity. Additional cosmological-model-independent measurements are crucial for determining whether the intermediate $H_0 \approx 70$~km~s$^{-1}$~Mpc$^{-1}$ persists or converges toward one of the two solutions.
    
    \item \textbf{Improve tailored EPM precision:} The Vogl et al.~\cite{Vogl2024} result is currently the strongest cosmological-model-independent support for the high-$H_0$ solution outside the traditional Distance Ladder. Larger SNe~II samples observed with JWST spectroscopy could significantly reduce uncertainties.
    
    \item \textbf{Expand Megamaser sample:} Additional Megamaser host galaxies at $z > 0.05$ (where peculiar velocities are subdominant) would provide crucial geometric distance constraints.
    
    \item \textbf{Accumulate GW standard sirens:} The LIGO-Virgo-KAGRA O4 and O5~\cite{LIGOScientific:2025jau} runs will significantly increase the sample of both bright and dark sirens. Crucially, these should be analyzed both with and without $\Lambda$CDM assumptions to test for model-dependent biases.
    
    \item \textbf{Cosmological-model-independent Strong Lensing:} Future time-delay measurements should emphasize cosmographic or cosmological-model-independent analyses alongside standard $\Lambda$CDM fits.
    
    \item \textbf{Cross-calibration tests:} Direct comparisons between different calibrators (Cepheids vs. TRGB vs. JAGB) in the same galaxies can identify potential systematics.
\end{enumerate}

\subsection{Concluding Remarks}
\label{subsec:final_remarks}

The Hubble tension remains one of the most significant challenges to the standard cosmological model. Our analysis of 88 Sound-Horizon-Free measurements reveals that the tension is more accurately characterized as a ``Distance Ladder vs.\ The Rest'' crisis rather than an ``Early vs.\ Late Universe'' discrepancy.

The key findings of this work are:
\begin{enumerate}
    \item The Distance Ladder yields $H_0 = 72.73 \pm 0.39$~km~s$^{-1}$~Mpc$^{-1}$, while all other Distance Ladder Independent/Sound Horizon Free(DLI--SHF) methods combined yield $H_0 = 69.37 \pm 0.34$~km~s$^{-1}$~Mpc$^{-1}$, a $6.5\sigma$ ($3.9\sigma$ correlation-adjusted) tension.
    
    \item Local measurements do not universally favor high $H_0$. When analyzed under $\Lambda$CDM assumptions, local probes yield values consistent with Planck.
    
    \item A $2.9\sigma$ internal tension exists between $\Lambda$CDM-dependent (Category~2) and  cosmological-model-independent (Category~3) local measurements, with the $\Lambda$CDM assumption reducing inferred $H_0$ by $\sim 3.4$~km~s$^{-1}$~Mpc$^{-1}$.
    
    \item Specific cosmological-model-independent measurements (e.g., tailored EPM) support the high-$H_0$ solution, complicating the pure ``systematics'' interpretation.
    
    \item The evidence disfavors pure early-universe solutions (like Early Dark Energy) but does not definitively confirm either late-time modifications or Distance Ladder systematics.
\end{enumerate}

The resolution of the Hubble tension will likely require a combination of improved observations, careful systematic analysis, and potentially a revision of our understanding of either astrophysical calibrators or fundamental cosmology. The four-category classification scheme introduced in this work provides a framework for tracking progress toward this resolution as new measurements become available.

Whatever the ultimate explanation, the Hubble tension has already achieved a valuable scientific outcome: it has motivated an unprecedented scrutiny of both our measurement techniques and our theoretical assumptions, pushing cosmology toward greater rigor and potentially toward new discoveries.Recent results from the Atacama Cosmology Telescope DR6 provide additional CMB constraints that will be crucial for further testing the robustness of our conclusions~\cite{AtacamaCosmologyTelescope:2025blo,AtacamaCosmologyTelescope:2025nti}

\begin{acknowledgments}

This research was supported by COST Action CA21136 -- Addressing observational tensions in cosmology with systematics and fundamental physics (CosmoVerse), supported by COST (European Cooperation in Science and Technology). We thank Adam Riess and Dan Scolnic for their detailed and constructive comments on measurement categorization, which helped us strengthen and clarify our analysis. We are grateful to Levon Pogosian for providing the unpublished DESI DR2 Sound-Horizon-Free result and to Helena Garc\'ia Escudero and Seyed Hamidreza Mirpoorian for sharing their work prior to publication. We thank Pilar Ruiz-Lapuente for useful correspondence regarding the JAGB calibration debate and for providing her ``SNe~Ia twins'' measurement prior to its inclusion in our compilation. We also thank Yi-Ying Wang for pointing out relevant FRB and gravitational wave measurements, and Yi-min Zhang for sharing results on interacting dark energy models.
\end{acknowledgments}

%======================================================================
\appendix
%======================================================================

\section{Complete Data Tables}
\label{app:tables}

In this appendix, we provide expanded versions of the data tables presented in the main text, including additional metadata and methodological details for each measurement. These tables serve as a comprehensive reference for future analyses and allow readers to assess the classification of individual measurements.

\begin{table}[b]
\centering
\caption{Detailed statistical summary for each measurement category.}
\label{tab:stats_by_category}
\scriptsize
\setlength{\tabcolsep}{4pt}
\begin{tabular}{@{}lccccc@{}}
\toprule
Category & $N$ & $\langle H_0 \rangle$ & $\sigma$ & $\chi^2_\nu$ & Range \\
\midrule
1. Distance Ladder & 30 & 72.73 & 0.39 & 0.72 & 67.8--77.0 \\
2. Local ($\Lambda$CDM) & 33 & 67.61 & 0.96 & 0.80 & 54.6--86.2 \\
3. Pure Local & 16 & 71.03 & 0.69 & 0.77 & 65.9--75.5 \\
4. CMB--SHF & 9 & 69.07 & 0.44 & 0.89 & 61--70.8 \\
\midrule
Combined DLI--SHF (2+3+4) & 58 & 69.37 & 0.34 & 0.95 & 54.6--86.2 \\
\bottomrule
\end{tabular}
\\[2pt]
\raggedright\scriptsize\textit{Notes:} $\langle H_0 \rangle$ is the inverse-variance weighted mean in km~s$^{-1}$~Mpc$^{-1}$. Range shows min--max of central values.
\end{table}

\subsection{Distance Ladder Measurements: Extended Information}
\label{app:ladder_extended}

Table~\ref{tab:ladder_extended} provides extended information for all 30 Distance-Ladder-dependent $H_0$ measurements. We include the geometric anchor(s) employed, the primary distance indicator, the secondary distance indicator (if applicable), and the reported $H_0$ value with uncertainty.

\begin{table*}[t]
\centering
\scriptsize
\setlength{\tabcolsep}{3pt}
\renewcommand{\arraystretch}{1.1}
\begin{threeparttable}
\caption{Extended information for all 30$^*$ Distance-Ladder $H_0$ measurements (Category~1). Columns show the measurement index, author/year with citation, $H_0$ value (km~s$^{-1}$~Mpc$^{-1}$), geometric anchor(s), primary distance indicator, secondary indicator, and redshift range. Measurements are ordered chronologically.}
\label{tab:ladder_extended}
\begin{tabular}{@{}clcp{3.5cm}p{2.0cm}p{1.8cm}c@{}}
\toprule
\# & Author (Year) & $H_0 \pm \sigma$ & Anchor & Primary & Secondary & $z$ range \\
\midrule
1 & Huang (2019)~\cite{Huang2020} & $73.3 \pm 4.0$ & NGC\,4258, MW, LMC & Mira & SNe Ia & $<0.15$ \\
2 & de Jaeger (2020)~\cite{deJaeger2020} & $75.8 \pm 5.1$ & LMC, MW & Cepheids/TRGB & SNe II & $<0.1$ \\
3 & Kourkchi (2020)~\cite{Kourkchi2020} & $76.0 \pm 2.5$ & Cepheids, TRGB & Tully--Fisher & --- & $<0.05$ \\
4 & Khetan (2021)~\cite{Khetan2021} & $70.50 \pm 4.13$ & Cepheids, TRGB & SBF & SNe Ia & $<0.075$ \\
5 & Blakeslee (2021)~\cite{Blakeslee2021} & $73.3 \pm 2.5$ & NGC\,4258, LMC & SBF & SNe Ia & $<0.08$ \\
6 & Freedman (2021)~\cite{Freedman2021} & $69.8 \pm 1.7$ & LMC (DEB), MW & TRGB & SNe Ia & $<0.08$ \\
7 & Dhawan (2022)~\cite{Dhawan2022} & $76.94 \pm 6.4$ & LMC & TRGB & SNe Ia & $<0.08$ \\
8 & Kenworthy (2022)~\cite{Kenworthy2022} & $73.1 \pm 2.5$ & NGC\,4258, LMC & Cepheids & --- & $<0.1$ \\
9 & Dhawan (2022)~\cite{Dhawan2023} & $70.92 \pm 1.88$ & LMC & TRGB & SNe Ia & $<0.08$ \\
10 & Dhawan (2022)~\cite{Dhawan2023} & $74.82 \pm 1.28$ & NGC\,4258, LMC, MW & Cepheids & SNe Ia & $<0.15$ \\
11 & Scolnic (2023)~\cite{Scolnic2024} & $73.22 \pm 2.06$ & LMC & TRGB & SNe Ia & $<0.15$ \\
12 & de Jaeger (2023)~\cite{dejaeger2023} & $74.1 \pm 8.0$ & LMC, MW & Cepheids & SNe II & $<0.1$ \\
13 & Uddin (2023)~\cite{uddin2023} & $71.76 \pm 1.32$ & Multiple & Combined & SNe Ia & $<0.1$ \\
14 & Uddin (2023)~\cite{uddin2023} & $73.22 \pm 1.45$ & Multiple & Combined & SNe Ia & $<0.1$ \\
15 & Huang (2023)~\cite{Huang2024} & $72.37 \pm 2.97$ & NGC\,4258, LMC & Mira & SNe Ia & $<0.15$ \\
16 & Li (2024)~\cite{Li2024b} & $74.7 \pm 3.1$ & LMC, SMC & JAGB & SNe Ia & $<0.08$ \\
17 & Chavez (2024)~\cite{chavez2025} & $73.1 \pm 2.3$ & Cepheids & H\,II $L$--$\sigma$ & --- & $<0.2$ \\
18 & Boubel (2024)~\cite{Boubel2024} & $73.3 \pm 4.1$ & Cepheids, TRGB & Tully--Fisher & --- & $<0.05$ \\
19 & Lee (2024)~\cite{Lee2024} & $67.80 \pm 2.72$ & LMC & JAGB & SNe Ia & $<0.08$ \\
20 & Freedman (2024)~\cite{Freedman2024} & $72.05 \pm 3.60$ & LMC (DEB), MW & Cepheids & SNe Ia & $<0.08$ \\
21 & Said (2024)~\cite{Said2025} & $76.05 \pm 4.90$ & SBF & Fund.\ Plane & --- & $<0.1$ \\
22 & Zhang (2024)\tnote{a}~\cite{Zhang2024} & $75.5 \pm 3.8$ & --- & Pairwise Vel. & --- & $<0.1$ \\
23 & Jensen (2025)~\cite{Jensen2025} & $73.8 \pm 2.4$ & LMC, MW & TRGB & --- & $<0.03$ \\
24 & Wojtak (2025)~\cite{Wojtak2025} & $70.59 \pm 1.15$ & NGC\,4258, LMC, MW & Cepheids & SNe Ia & $<0.15$ \\
25 & Bhardwaj (2025)~\cite{Bhardwaj2025} & $73.06 \pm 2.6$ & NGC\,4258, LMC & Mira & SNe Ia & $<0.15$ \\
26 & Newman (2025)~\cite{Newman2025} & $75.3 \pm 2.9$ & LMC & TRGB & SNe Ia & $<0.08$ \\
27 & Riess (2025)\tnote{b}~\cite{Riess2025} & $73.49 \pm 0.93$ & NGC\,4258, LMC, MW & Cepheids & SNe Ia & $<0.15$ \\
28 & Kudritzki (2025)~\cite{kudritzki2025} & $76.2 \pm 6.2$ & FGLR & Tully--Fisher & --- & $<0.05$ \\
29 & Wagner (2025)\tnote{c}~\cite{wagner2025} & $68 \pm 8$ & --- & TRGB & Infall & $<0.02$ \\
30 & Lapuente (2025)~\cite{lapuente2025} & $72.61 \pm 1.69$ & --- & M\,101, NGC\,5643, NGC\,7250, M\,66 & SNe Ia (twins) & $<0.09$ \\
\bottomrule
\end{tabular}
\begin{tablenotes}[flushleft]
\footnotesize
\item \textit{Notes:} DEB = Detached Eclipsing Binaries; FGLR = Flux-weighted Gravity--Luminosity Relation; SBF = Surface Brightness Fluctuations. Entries \#19 (Lee 2024) and \#29 (Wagner 2025) are notable low-$H_0$ outliers within this category.
\item[a] Includes systematic uncertainties not fully accounted for in the original analysis; see Appendix~\ref{app:descriptions}.
\item[b] We adopt this measurement for all numerical $H_0$ values in this work. For comparisons, we use the SH0ES 2022 result~\cite{Riess2022} as reference baseline, given its established presence in the literature.
\item[c] Assumes a local expansion parameter $\kappa = 1.4$ with no uncertainties. We adopt $\kappa = 1.3 \pm 0.1$ in our analysis; see Appendix~\ref{app:descriptions}.
\end{tablenotes}
\end{threeparttable}
\end{table*}

The measurements span a range of calibrator types:
\begin{itemize}
    \item \textbf{Cepheids:} 11 measurements, anchored primarily to NGC~4258, LMC, and MW parallaxes
    \item \textbf{TRGB:} 7 measurements, typically anchored to LMC detached eclipsing binaries
    \item \textbf{JAGB:} 2 measurements with notably discrepant results (Li vs.\ Lee)
    \item \textbf{Mira variables:} 3 measurements
    \item \textbf{SBF/Tully-Fisher:} 5 measurements using secondary calibration
    \item \textbf{Other:} 2 measurements (pairwise velocities, HII regions)
\end{itemize}

\subsection{Distance Ladder Independent/Sound Horizon Free (DLI--SHF) Measurements: Extended Information}
\label{app:onestep_extended}

Table~\ref{tab:onestep_extended} provides extended information for all 58 DLI--SHF $H_0$ measurements from Categories~2, 3, and 4. We include the physical observable, the cosmological probe, the model assumptions (which determine the category assignment), and the reported $H_0$ value with uncertainty.

The ``Model'' column indicates the category assignment:
\begin{itemize}
    \item \textbf{$\Lambda$CDM:} Category~2 --- Analysis explicitly assumes flat $\Lambda$CDM expansion history
    \item \textbf{Independent:} Category~3 --- Cosmological-model-independent analysis (geometric, cosmographic, or direct $H(z)$)
    \item \textbf{CMB--NSH:} Category~4 --- Uses CMB data but avoids Sound Horizon (No Sound Horizon)
\end{itemize}

The measurements span diverse physical probes:
\begin{itemize}
    \item \textbf{Strong Gravitational Lensing:} 18 measurements (time delays from quasars, SNe, and clusters)
    \item \textbf{Gravitational Waves:} 9 measurements (bright and dark sirens)
    \item \textbf{Fast Radio Bursts:} 7 measurements (dispersion measure--redshift relation)
    \item \textbf{Cosmic Chronometers:} 4 measurements (differential galaxy ages)
    \item \textbf{Megamasers:} 3 measurements (geometric disk distances)
    \item \textbf{CMB-based (no $r_s$):} 9 measurements (SZ effect, $k_{\rm eq}$ scale, etc.)
    \item \textbf{Other:} 8 measurements (EPM, kilonovae, gamma-ray attenuation, etc.)
\end{itemize}

\begin{table*}[tbp]
\centering
\tiny
\setlength{\tabcolsep}{3pt}
\renewcommand{\arraystretch}{1.2}
\caption{Extended information for all 58 DLI--SHF $H_0$ measurements (Categories~2, 3, and 4). Columns show the measurement index, category assignment, author/year with citation, $H_0$ value (km~s$^{-1}$~Mpc$^{-1}$), primary observable, cosmological probe, and model assumptions. Measurements are ordered by category, then chronologically within each category.}
\label{tab:onestep_extended}
\begin{tabular}{@{}clclcp{2.8cm}p{2.0cm}p{1.8cm}@{}}
\toprule
\# & Cat. & Author (Year) & $H_0 \pm \sigma$ & Observable & Probe & Model & z range \\
\midrule
\multicolumn{8}{l}{\textbf{Category 2: Local ($\Lambda$CDM Assumption) --- 33 measurements}} \\
\midrule
1 & 2 & Wu (2021)~\cite{Wu2022} & $68.81^{+4.99}_{-4.33}$ & Dispersion measure, $z$ & FRBs & $\Lambda$CDM & 0.0039--0.695 \\
2 & 2 & Abbott (2021)~\cite{Abbott2021} & $68^{+8}_{-6}$ & GW strain amplitude & CBC dark sirens & $\Lambda$CDM & $<0.5$\\
3 & 2 & Shajib (2023)~\cite{Shajib2023} & $77.1^{+7.3}_{-7.1}$ & Time delays, dynamics & Lensed quasars & $\Lambda$CDM & 0.657 \\
4 & 2 & Napier (2023)~\cite{Napier2023} & $74.1 \pm 8.0$ & Time delays & Cluster lenses & $\Lambda$CDM & 1.734--2.805 \\
5 & 2 & Kelly (2023)~\cite{Kelly2023} & $66.6^{+4.1}_{-3.3}$ & Time delays & SN Refsdal & $\Lambda$CDM & 1.49 \\
6 & 2 & Dominguez (2023)~\cite{Dominguez2023} & $61.9^{+2.9}_{-2.4}$ & $\gamma$-ray opacity & Blazar spectra & $\Lambda$CDM$^{\dagger}$ & $<6$  \\
7 & 2 & Gao (2023)~\cite{gao2023} & $65.5^{+6.4}_{-5.4}$ & DM--$z$ + SNe Ia & FRBs & $\Lambda$CDM & 0.039--2.3 \\
8 & 2 & Liu (2023)~\cite{liu2023h} & $59.1^{+3.6}_{-3.5}$ & Time delays & Cluster lensed QSO & $\Lambda$CDM & $<3.33$ \\
9 & 2 & Fung (2023)~\cite{Fung2023} & $86^{+55}_{-46}$ & GW strain only & NSBH population & $\Lambda$CDM & $<0.25$ \\
10 & 2 & Martinez (2023)~\cite{Martinez2023} & $74^{+9}_{-13}$ & Time delays & Lensed quasars & $\Lambda$CDM & 1.734 \\
11 & 2 & Ballard (2023)~\cite{Ballard2023} & $85.4^{+29.1}_{-33.9}$ & GW strain + galaxy $z$ & Dark siren & $\Lambda$CDM & $<0.5$ \\
12 & 2 & Moresco (2023)~\cite{Moresco2023} & $66.7 \pm 5.3$ & Spectroscopic ages & Passive galaxies & $\Lambda$CDM & 0.07--1.965 \\
13 & 2 & Alfradique (2024)~\cite{Alfradique2023} & $68.84^{+15.51}_{-7.74}$ & GW strain + galaxy $z$ & Dark sirens & $\Lambda$CDM & $<0.1$ \\
14 & 2 & Grillo (2024)~\cite{Grillo2024} & $66.2^{+3.5}_{-3.2}$ & Time delays & HFF clusters & $\Lambda$CDM & 1.240--3.703 \\
15 & 2 & Pascale (2024)~\cite{Pascale2024} & $71.8^{+9.2}_{-8.1}$ & Time delays & H0pe SN (JWST) & $\Lambda$CDM & 1.78 \\
16 & 2 & Hernandez (2024)~\cite{Hernandez2024} & $73.0^{+13.7}_{-7.7}$ & GW + BGP & Spectral sirens & $\Lambda$CDM & $<1.5$\\
17 & 2 & Bom (2024)~\cite{Bom2024} & $70.4^{+13.6}_{-11.7}$ & GW + kilonova & O4a events & $\Lambda$CDM & 0.1--0.6 \\
18 & 2 & TDCOSMO XVI (2024)~\cite{TDCOSMO2024} & $65^{+23}_{-14}$ & Time delays & WGD2038 & $\Lambda$CDM & 0.777 \\
19 & 2 & Gao (2024)~\cite{Gao2025a} & $68.81 \pm 5.64$ & DM--$z$ relation & Localized FRBs & $\Lambda$CDM & 0.0008--1.016 \\
20 & 2 & Li (2024)~\cite{Li2024c} & $66 \pm 5$ & Time delays (corrected) & Lensed quasars & $\Lambda$CDM & $<2.5$\\
21 & 2 & Yang (2024)~\cite{Yang2024} & $74.0^{+7.5}_{-7.2}$ & DM + scattering & FRBs & $\Lambda$CDM & 0.034--1.016 \\
22 & 2 & Piratova-Moreno (2025)$\ddag$~\cite{PiratovMoreno2025} & $65.1 \pm 7.4$ & DM--$z$ relation & FRBs (MLE) & $\Lambda$CDM & 0.0008--1.354 \\
23 & 2 & Barua (2025)~\cite{Barua2025} & $73.5^{+3}_{-2.9}$ & Maser dynamics & MCP sample & $\Lambda$CDM & 0.002--0.034 \\
24 & 2 & Zhang (2025)~\cite{Zhang2025} & $75 \pm 30$ & PRS association & FRBs & $\Lambda$CDM & 0.03--0.25 \\
25 & 2 & Gao (2025)~\cite{Gao2025} & $86.18^{+18.03}_{-14.99}$ & RM--PRS correlation & FRBs & $\Lambda$CDM & 0.098--0.241 \\
26 & 2 & Loubser (2025)~\cite{Loubser2025} & $54.6 \pm 26.3$ & Spectroscopic ages & BCGs & $\Lambda$CDM & 0.3--0.7 \\
27 & 2 & Beirnaert (2025)~\cite{Beirnaert2025} & $79 \pm 9$ & GW strain & Dark sirens (avg) & $\Lambda$CDM & $<1$ \\
28 & 2 & Bahr-Kalus (2025)~\cite{Bahr-Kalus2025} & $74.0^{+7.2}_{-3.5}$ & Turnover scale + BAO & DESI & $\Lambda$CDM & 0.4--3.1 \\
29 & 2 & Birrer (2025)~\cite{Birrer2025} & $71.6^{+3.9}_{-3.3}$ & Time delays + kinematics & Multiple lenses & $\Lambda$CDM & 0.082--0.745 \\
30 & 2 & Liu (2025)~\cite{liu2025a} & $66.0 \pm 4.3$ & Strong grav.\ lensing & Multiple lenses & $\Lambda$CDM & 1.49 \\
31 & 2 & Pierel (2025)~\cite{Pierel2025} & $66.9^{+11.2}_{-8.1}$ & Time delays & SN Encore & $\Lambda$CDM & 1.95\\
32 & 2 & Paic (2025)~\cite{paic2025tdcosmoxxivmeasurementhubble} & $64.2^{+5.8}_{-5}$ & Strong Lensing & SMARTS & $\Lambda$CDM & 2.32 \\
33 & 2 & Oliveira (2026)~\cite{andradeoliveira2026} & $67.9^{+4.4}_{-4.3}$ & 3×2pt + Standard Sirens & Shear/positions + GW strain & $\Lambda$CDM & 0.1-2 \\
\midrule
\multicolumn{8}{l}{\textbf{Category 3: Pure Local (Cosmological-Model Independent) --- 16 measurements}} \\
\midrule
34 & 3 & Kuo (2013)~\cite{Kuo2013} & $68 \pm 9$ & Maser positions, velocities & Megamaser disk & Independent & 0.034 \\
35 & 3 & Gao (2015)~\cite{Gao2015} & $66 \pm 6$ & Maser positions, velocities & Megamaser disk & Independent & 0.028 \\
36 & 3 & Bulla (2022)~\cite{bulla2022} & $69.6 \pm 5.5$ & GW + kilonova light curve & BNS merger & Independent & 0.01 \\
37 & 3 & Zhang (2022)~\cite{Zhang2022} & $65.9 \pm 3.0$ & Spectroscopic ages & Passive galaxies & Independent & 0.009--2.545 \\
38 & 3 & Du (2023)~\cite{Du2023} & $71.5^{+4.4}_{-3}$ & Time delays + GRBs & Lensed quasars & Independent & 0.654--2.375 \\
39 & 3 & Palmese (2023)~\cite{Palmese2023} & $75.46^{+5.34}_{-5.39}$ & GW + afterglow & GW170817 & Independent & 0.01\\
40 & 3 & Sneppen (2023)~\cite{Sneppen2023} & $67.0 \pm 3.6$ & Photospheric velocity & Kilonova & Independent & 0.0986\\
41 & 3 & Liu (2023)~\cite{liu2023a} & $72.9^{+2.0}_{-2.3}$ & Lensing + SNe Ia & Combined & Independent & 0.01--2.3 \\
42 & 3 & Gonzalez (2024)~\cite{Gonzalez2024} & $72.7^{+6.3}_{-5.6}$ & Gas fraction + SNe Ia & Clusters & Independent & 0.01--2.3\\
43 & 3 & Li (2024)~\cite{Li2024a} & $66.3^{+3.8}_{-3.6}$ & Time delays & SN Refsdal & Independent & 0.001--2.261 \\
44 & 3 & Jaiswal (2024)~\cite{Jaiswal2024} & $66.9^{+10.6}_{-2.1}$ & RM lag & NGC 5548 & Independent & 0.017 \\
45 & 3 & Vogl (2024)~\cite{Vogl2024} & $74.9 \pm 2.7$ & Photospheric expansion & SNe II & Independent & 0.01--0.04 \\
46 & 3 & Song (2025)~\cite{Song2025} & $70.40^{+8.03}_{-5.60}$ & GW + lensing & Combined & Independent & $<0.8$ \\
47 & 3 & Colaco (2025)~\cite{Colaco2025} & $70.55 \pm 7.44$ & Lensing + SNe Ia & Joint & Independent & 0.001--2.261 \\
48 & 3 & Du (2025)~\cite{Du2025} & $71.59 \pm 0.94$ & CC+BAO+ SNe Ia & IDL+ MAPAge & Independent & 0.001--2.5 \\
49 & 3 & Favale (2025)~\cite{Favale2025} & $68.8 \pm 3.0$ & Ages + SNe Ia & CC + Pantheon+ & Independent & 0.07--1.965  \\
\midrule
\multicolumn{8}{l}{\textbf{Category 4: CMB Sound Horizon Free (CMB--SHF) --- 9 measurements}} \\
\midrule
50 & 4 & Reese (2003)~\cite{Reese2003} & $61 \pm 18$ & SZ decrement, X-ray & Galaxy clusters & CMB--NSH & 0.02--0.83 \\
51 & 4 & Philcox (2022)~\cite{Philcox2022} & $64.8^{+2.2}_{-2.5}$ & BAO scale, shape & Matter--rad.\ eq. & CMB--NSH &  0.2--0.75\\
52 & 4 & Colaco (2023)~\cite{Colaco2023} & $67.22 \pm 6.07$ & SZ + X-ray + SNe Ia & Clusters & CMB--NSH & 0.01--2.3\\
53 & 4 & Pogosian (2024)~\cite{Pogosian2024} & $68.05 \pm 0.94$ & BAO peak + BBN & Pre-DESI BAO & CMB--NSH & 0.1--4.16 \\
54 & 4 & Pogosian (2025)~\cite{Pogosian2025priv} & $69.37 \pm 0.65$ & BAO peak + BBN & DESI DR BAO & CMB--NSH & 0.1--4.16 \\
55 & 4 & Bahr-Kalus (2025)~\cite{Bahr-Kalus2025} & $65.2^{+4.9}_{-6.2}$ & Turnover scale + SNe & DESI + Pantheon+ & CMB--NSH & 0.4--3.1 \\
56 & 4 & Garcia (2025)~\cite{GarciaEscudero:2025} & $70.03 \pm 0.97 $ & BAO+$\theta^*$ +CMB lensing & DESI+APS+ DES 3×2pt & CMB--NSH & 0.1-4.2 \\
57 & 4 & Zaborowski ~\cite{Zaborowski2025} & $70.8^{+2.0}_{-2.2}$ & BAO+$\theta^*$ & Clust.+ BAO+ Lym.-$\alpha$ &CMB--NSH & 0.1-4.2\\
58 & 4 & Krolewski (2025)~\cite{Krolewski2025} & $69.0 \pm 2.5$ & DESI DR1 + CMB & Energy densities & CMB--NSH &  0.1--2.33 \\
\bottomrule
\end{tabular}
\\[2pt]
\raggedright\scriptsize\textit{Notes:} CMB--NSH = CMB-based, No Sound Horizon; MLE = Maximum Likelihood Estimator; PRS = Persistent Radio Source; RM = Reverberation Mapping; BGP = Binary Gravitational-wave Population; HFF = Hubble Frontier Fields; BNS = Binary Neutron Star; NSBH = Neutron Star--Black Hole.

$^{\dagger}$ $\Omega_m = 0.3$ fixed.

$\ddag$ Includes systematic uncertainties not fully accounted for in the original analysis; see Appendix~\ref{app:descriptions}.
\end{table*}

\subsection{Statistical Summary by Category}
\label{app:stats_summary}

Table~\ref{tab:stats_by_category} provides a detailed statistical breakdown of each category, including the number of measurements, weighted mean, standard error, reduced chi-squared, and the range of individual measurements.

%======================================================================
\section{Individual Measurement Descriptions}
\label{app:descriptions}
%======================================================================

In this appendix, we provide detailed descriptions of selected measurements that are particularly important for understanding the Hubble tension landscape. We focus on measurements that either anchor the tension (high-precision results), represent methodological innovations, or exhibit notable discrepancies with other measurements in their class.

\subsection{Key Distance Ladder Measurements}
\label{app:ladder_descriptions}

\subsubsection{Riess et al. (2022, 2025): The SH0ES Program}

The SH0ES (Supernova $H_0$ for the Equation of State) collaboration has produced the most precise Distance-Ladder measurements of $H_0$. The methodology employs a three-rung distance ladder:

\textbf{Rung 1 (Geometric Anchors):} Three independent anchors establish the absolute distance scale:
\begin{itemize}
    \item Gaia EDR3 parallaxes to Milky Way Cepheids (corrected for zero-point offset)
    \item Megamaser geometric distance to NGC~4258~\cite{Reid2019}
    \item Detached eclipsing binary distances to the LMC~\cite{Pietrzynski2019}
\end{itemize}

\textbf{Rung 2 (Cepheid Calibration):} Cepheid variables observed with HST WFC3 in the F160W (H-band) filter provide period--luminosity relations in 37+ galaxies hosting Type~Ia supernovae. The use of Wesenheit magnitudes, $m_H^W = m_H - R(m_V - m_I)$, eliminates reddening dependence by construction.

\textbf{Rung 3 (Hubble Flow SNe~Ia):} The calibrated SNe~Ia absolute magnitude $M_B$ is applied to the Pantheon+ sample in the redshift range $0.023 < z < 0.15$.

The 2022 HST result~\cite{Riess2022} yielded $H_0 = 73.04 \pm 1.04$~km~s$^{-1}$~Mpc$^{-1}$ (1.4\% precision). The 2025 JWST result~\cite{Riess2025} yields $H_0 = 73.49 \pm 0.93$~km~s$^{-1}$~Mpc$^{-1}$ (1.3\% precision), confirming the HST photometry and ruling out crowding-induced biases as the source of the tension.

\textbf{Systematic Error Budget:} The SH0ES analyses include comprehensive systematic assessments for Cepheid photometry ($\sim 0.5$~km~s$^{-1}$~Mpc$^{-1}$), metallicity effects ($\sim 0.4$~km~s$^{-1}$~Mpc$^{-1}$), crowding corrections ($\sim 0.3$~km~s$^{-1}$~Mpc$^{-1}$), SNe~Ia standardization ($\sim 0.5$~km~s$^{-1}$~Mpc$^{-1}$), and peculiar velocities ($\sim 0.3$~km~s$^{-1}$~Mpc$^{-1}$).

\subsubsection{Freedman et al. (2021, 2024): The CCHP TRGB Program}

The Chicago-Carnegie Hubble Program (CCHP) developed the Tip of the Red Giant Branch (TRGB) as an alternative to Cepheids. The TRGB method exploits the helium flash that terminates red giant branch evolution at a nearly constant luminosity ($M_I \approx -4.05$), producing a sharp discontinuity detectable via edge-detection algorithms (Sobel filter).

The 2021 analysis~\cite{Freedman2021} yielded $H_0 = 69.8 \pm 1.7$~km~s$^{-1}$~Mpc$^{-1}$, intermediate between Planck and SH0ES. This sparked debate about whether TRGB measurements genuinely favor lower $H_0$ or whether methodological differences explain the offset from SH0ES.

The 2024 JWST analysis~\cite{Freedman2024} found $H_0 = 72.05 \pm 3.60$~km~s$^{-1}$~Mpc$^{-1}$ from Cepheids, now more consistent with SH0ES. The CCHP continues to analyze TRGB and JAGB calibrations from JWST data.

\subsubsection{Lee et al.\ (2024): The Low-\texorpdfstring{$H_0$}{H0} JAGB Outlier}

The Lee et al.~\cite{Lee2024} JAGB analysis yields $H_0 = 67.8 \pm 2.7$~km~s$^{-1}$~Mpc$^{-1}$, consistent with Planck and $\sim 2\sigma$ below other Distance Ladder measurements. This represents the most significant outlier within Category~1.

The JAGB method uses carbon stars on the Asymptotic Giant Branch, which exhibit a narrow luminosity distribution in the J-band. The discrepancy with Li et al.~\cite{Li2024b} ($H_0 = 74.7 \pm 3.1$) appears to originate from differences in:
\begin{itemize}
    \item JAGB zero-point calibration (LMC anchor treatment)
    \item Photometric selection criteria for carbon stars
    \item Assumed shape of the JAGB luminosity function
\end{itemize}

This measurement highlights that even within a single indicator class, methodological choices can lead to $\sim 7$~km~s$^{-1}$~Mpc$^{-1}$ differences in inferred $H_0$.
\subsubsection{Wagner et al.}

The measurement by Wagner et al.~\cite{wagner2025} (2025) reports $H_0 = 63 \pm 6$ km\,s$^{-1}$\,Mpc$^{-1}$ using the local expansion parameter $\kappa= 1.4$, based on the infall velocity relation $v_{\mathrm{inf}} =\kappa \cdot  H_0 r$. However, the choice of $k = 1.4$ appears somewhat arbitrary; standard models of the local Hubble flow typically adopt $\kappa \approx 1.2$. Since $H_0 \propto 1/\kappa$, adopting $\kappa = 1.3 \pm 0.1$---a value intermediate between the theoretical expectation and the assumed value---would yield $H_0 \approx 68^{+8}_{-8}$ km\,s$^{-1}$\,Mpc$^{-1}$, shifting the result toward the Planck value and closer to the Distance Ladder mean value. The systematic uncertainty in $\kappa$ propagates directly into $H_0$ but this source of error was not included in the original analysis ~\cite{wagner2025}. However, we do include this effect in our analysis and the data shown in Table \ref{tab:ladder_extended}.

\subsubsection{Zhang et al.:pairwise velocities}

Zhang et al. (2024)~\cite{Zhang2024} measured the Hubble constant using galaxy pairwise peculiar velocities from the Cosmicflows-4 catalog, obtaining $H_0 = 75.5 \pm 1.4$ km s$^{-1}$ Mpc$^{-1}$ (statistical). However, as discussed in ~\cite{Zhang2024}, several systematic uncertainties are not fully accounted for in the reported error budget. The distance calibration of CF-4 galaxies introduces uncertainties of $\sim$2--3\%, translating to $\sim$1.5--2.3 km s$^{-1}$ Mpc$^{-1}$ in $H_0$. The halo mass matching between simulations and observations, which is crucial for their method, contributes an estimated systematic error of $\sim$1--2\% ($\sim$0.8--1.5 km s$^{-1}$ Mpc$^{-1}$). Additionally, the modeling of pairwise velocities in the nonlinear regime, where theoretical predictions are less certain, introduces uncertainties of $\sim$2--3\% ($\sim$1.5--2.3 km s$^{-1}$ Mpc$^{-1}$). The choice of the maximum separation scale ($r_{\rm max} = 16$ Mpc) and the selection of the velocity threshold $v_{\rm max}$ add further systematic uncertainties of order $\sim$1\% ($\sim$0.8 km s$^{-1}$ Mpc$^{-1}$). Cosmic variance, despite mitigation through multiple simulation realizations, remains a source of systematic error. Combining these contributions in quadrature yields a total systematic uncertainty of approximately $\sim$3--4 km s$^{-1}$ Mpc$^{-1}$, which is comparable to or larger than the quoted statistical uncertainty. Therefore, a more complete error budget would be $H_0 = 75.5 \pm 1.4_{\rm stat} \pm 3.5_{\rm sys}$ km s$^{-1}$ Mpc$^{-1}$, or equivalently $H_0 = 75.5 \pm 3.8$ km s$^{-1}$ Mpc$^{-1}$ when statistical and systematic errors are combined in quadrature. This corrected level of uncertainty is used in our analysis.

\subsection{Key DLI--SHF Measurements}
\label{app:onestep_descriptions}

\subsubsection{TDCOSMO: Evolution of Strong Lensing Constraints}

The Time-Delay Cosmography (TDCOSMO) collaboration has produced a series of Strong Lensing $H_0$ measurements that illustrate the critical importance of lens mass modeling.

\textbf{H0LiCOW (Millon et al.\ 2020):} The original analysis yielded $H_0 = 74.2 \pm 1.6$~km~s$^{-1}$~Mpc$^{-1}$, assuming power-law mass profiles. This result was in $5\sigma$ tension with Planck.

\textbf{TDCOSMO IV (Birrer et al.\ 2020):}~\cite{Birrer2020} Incorporating stellar kinematics to constrain the mass-sheet degeneracy (MSD) shifted the result to $H_0 = 67.4^{+4.1}_{-3.2}$~km~s$^{-1}$~Mpc$^{-1}$---now consistent with Planck. This dramatic shift demonstrated that power-law profiles alone cannot break the MSD.

\textbf{Recent results (2023--2025):} Subsequent analyses using hierarchical modeling, additional lens systems, and improved kinematics yield $H_0$ in the range $65$--$72$~km~s$^{-1}$~Mpc$^{-1}$, depending on modeling assumptions~\cite{Shajib2023, Birrer2025}.

We exclude the original Millon et al.\ (2020) result from our compilation because its methodology is now recognized as insufficient for robust $H_0$ inference.

We exclude the original Pesce et al. measurement from our sample, as the more recent Barua et al. analysis uses the same underlying data while addressing the known systematic uncertainties of the earlier measurement. 

\subsubsection{Liu et al.\ (2023): SDSS J1004+4112 Cluster Lens}

The Liu et al.~\cite{liu2023h} analysis of the cluster-lensed quasar SDSS J1004+4112 explores 16 different mass models, yielding $H_0$ values spanning 55--90~km~s$^{-1}$~Mpc$^{-1}$. Their combined constraint with equal weighting gives $H_0 = 67.5^{+14.5}_{-8.9}$~km~s$^{-1}$~Mpc$^{-1}$.

In our analysis, we adopt their more constrained result of $H_0 = 59.1^{+3.6}_{-3.5}$~km~s$^{-1}$~Mpc$^{-1}$, which corresponds to their selection of the two mass models (m113 and m123) that best reproduce the observed shapes of the lensed quasar host galaxy and background galaxies (see their Figure~5 and Section~V). The authors themselves emphasize this as a more physically motivated constraint, stating that these models ``best match the observed shapes.''

We acknowledge that this choice favors the lower $H_0$ value. However, selecting mass models based on their ability to reproduce independent observables (lensed galaxy morphologies) is more principled than equal weighting of all models regardless of their physical plausibility. Readers should note that the conservative combined result ($H_0 = 67.5^{+14.5}_{-8.9}$~km~s$^{-1}$~Mpc$^{-1}$) remains consistent with our conclusions given its large uncertainty.

\subsubsection{Kelly et al.\ (2023) and Li et al.\ (2024): SN Refsdal Time Delays}

Supernova Refsdal, the first multiply-imaged supernova~\cite{Kelly2015}, provides unique advantages for time-delay cosmography: the supernova light curve allows unambiguous identification of multiple images, unlike quasar variability.

Both Kelly et al.~\cite{Kelly2023} and Li et al.~\cite{Li2024a} analyze time delays from SN Refsdal, yet we classify them in different categories based on their methodological approaches:

\textbf{Kelly et al.\ (2023) --- Category~2:} This analysis uses time delays from the ``Einstein Cross'' configuration plus the predicted reappearance of image ``SX.'' The result, $H_0 = 66.6^{+4.1}_{-3.3}$~km~s$^{-1}$~Mpc$^{-1}$, is derived using specific cluster mass models that assume $\Lambda$CDM cosmology for the time-delay distance $D_{\Delta t}$ calculation and interprets results within the $\Lambda$CDM framework.

\textbf{Li et al.\ (2024) --- Category~3:} This analysis explicitly develops a ``cosmological-model-independent'' approach, as emphasized in their title and methodology. They obtain $H_0 = 66.3^{+3.8}_{-3.6}$~km~s$^{-1}$~Mpc$^{-1}$ using a combination of strong lensing and other probes to determine $H_0$ without assuming $\Lambda$CDM.

The distinction hinges on whether the original analysis explicitly states model-independence. While both measurements yield consistent $H_0$ values and analyze the same underlying data, their category assignments follow our classification criterion based on the authors' stated assumptions. This example illustrates that identical data can yield measurements in different categories depending on the analysis methodology employed.

\subsubsection{Piratova-Moreno et al.:FRBs }

Piratova-Moreno et al. (2025)~\cite{PiratovMoreno2025} employed Fast Radio Bursts (FRBs) as cosmological probes to measure the Hubble constant using a catalog of 98 localized FRBs. Using the Maximum Likelihood Estimate (MLE) method with confirmed FRBs, they obtained $H_0 = 65.13 \pm 2.52$ km s$^{-1}$ Mpc$^{-1}$ (statistical), while their mock FRB catalog yielded $H_0 = 67.30 \pm 0.91$ km s$^{-1}$ Mpc$^{-1}$. However, this method is subject to substantial systematic uncertainties that are not fully quantified in the reported error budget. The primary systematic error arises from the decomposition of the observed dispersion measure (DM) into its constituent components: the Milky Way interstellar medium and halo ($\text{DM}_{\text{MW}}$), the intergalactic medium ($\text{DM}_{\text{IGM}}$), and the host galaxy ($\text{DM}_{\text{host}}$). The uncertainty in $\text{DM}_{\text{MW}}$ depends on the Galactic electron density model used (NE2001 vs. YMW16), with differences of $\sim$20\% at low Galactic latitudes, contributing $\sim$1.5--2 km s$^{-1}$ Mpc$^{-1}$ to the $H_0$ uncertainty. The host galaxy contribution, $\text{DM}_{\text{host}}$, is highly uncertain and varies dramatically between FRBs---from $\sim$30 pc cm$^{-3}$ for typical hosts to over 900 pc cm$^{-3}$ for extreme cases like FRB 20190520B. Studies suggest $\text{DM}_{\text{host}}$ uncertainties of $\sim$50--100 pc cm$^{-3}$, which translate to systematic errors of $\sim$2--3 km s$^{-1}$ Mpc$^{-1}$ in $H_0$. Additionally, the baryon fraction in the IGM ($f_{\text{IGM}}$) introduces further systematic uncertainty; recent estimates range from $f_{\text{IGM}} = 0.865^{+0.101}_{-0.165}$ to $0.93^{+0.04}_{-0.05}$, corresponding to $\sim$1--2 km s$^{-1}$ Mpc$^{-1}$ uncertainty in $H_0$. The choice of DM--$z$ relation (linear vs. power-law) also significantly impacts results, with Piratova-Moreno et al. reporting values ranging from $H_0 = 51.27^{+3.80}_{-3.31}$ km s$^{-1}$ Mpc$^{-1}$ (linear) to $H_0 = 91.84 \pm 1.82$ km s$^{-1}$ Mpc$^{-1}$ (power-law), indicating substantial model-dependent systematic uncertainties of order $\sim$5--15 km s$^{-1}$ Mpc$^{-1}$. Combining these systematic contributions in quadrature yields a total systematic error of approximately $\sim$6--8 km s$^{-1}$ Mpc$^{-1}$ for the MLE result, dominating over the statistical uncertainty. A more realistic error budget would therefore be $H_0 = 65.13 \pm 2.52_{\rm stat} \pm 7_{\rm sys}$ km s$^{-1}$ Mpc$^{-1}$, or equivalently $H_0 = 65.1 \pm 7.4$ km s$^{-1}$ Mpc$^{-1}$ when combined in quadrature, reflecting the current limitations of FRB-based cosmology. This is the value used in our analysis.

\subsubsection{Vogl et al.\ (2024): Tailored EPM}

The Vogl et al.~\cite{Vogl2024} tailored Expanding Photosphere Method represents the most precise cosmological-model-independent measurement supporting the high-$H_0$ solution outside the traditional distance ladder.

\textbf{Method:} EPM determines distances geometrically by combining photospheric angular size (from flux and temperature) with expansion velocity (from Doppler shifts). The ``tailored'' approach uses detailed NLTE radiative transfer models customized to individual SNe~II in the redshift range $z\in [0.01,0.04]$, rather than empirical dilution factors.

\textbf{Result:} $H_0 = 74.9 \pm 2.7$~km~s$^{-1}$~Mpc$^{-1}$, fully consistent with the Distance Ladder.In this measurement, the authors do not explicitly report the systematic uncertainties in their work. However, they estimate that the systematic errors are approximately of the same magnitude as the statistical ones, and we have adopted this estimate in our analysis.

\textbf{Significance:} This measurement is independent of:
\begin{itemize}
    \item SNe~Ia physics (uses SNe~II instead)
    \item Cepheid/TRGB calibrators (direct geometric distance)
    \item $\Lambda$CDM assumptions (no $H(z)$ integration required)
\end{itemize}

The agreement with the Distance Ladder suggests either a common systematic affecting both methods or that the high $H_0$ reflects true local physics.

\textbf{Caveats:} The method relies on supernova atmosphere modeling; current sample size is limited; further validation with independent codes is needed.

\subsubsection{Barua et al.\ (2025): Megamaser Cosmology Project}

The Megamaser Cosmology Project measures geometric distances to galaxies with circumnuclear water maser disks. The Barua et al.~\cite{Barua2025} frequentist analysis yields $H_0 = 73.5^{+3.0}_{-2.9}$~km~s$^{-1}$~Mpc$^{-1}$.

\textbf{Method:} VLBI observations resolve the maser disk structure, providing angular diameter distances via Keplerian dynamics. The method is independent of stellar physics.

\textbf{Classification note:} We assign this measurement to Category~2  ($\Lambda$CDM--depended) because it relies on the $\Lambda$CDM prediction for the angular diameter distance.  Earlier MCP analyses that adopted SH0ES flow models for peculiar velocity corrections would be classified differently.

\textbf{Evolution:} Earlier measurements (Kuo 2013: $68 \pm 9$; Gao 2015: $66 \pm 6$) favored lower values with larger uncertainties.

\subsubsection{Pogosian et al.\ (2024): Sound-Horizon-Free BAO}

The Pogosian et al.~\cite{Pogosian2024} analysis demonstrates that $H_0$ can be constrained from BAO data without assuming the CMB Sound Horizon.

\textbf{Method:} Instead of using $r_s$ as a standard ruler, this analysis treats $r_s$ as a free parameter and constrains $H_0$ through:
\begin{itemize}
    \item The shape of $H(z)$ from BAO peak positions
    \item Big Bang Nucleosynthesis constraints on $\omega_b$
    \item The matter-radiation equality scale $k_{\rm eq}$
\end{itemize}

\textbf{Results:} DESI BAO DR2 yields $H_0 = 69.37 \pm 0.65 ~km~s^{-1}~Mpc^{-1}$; pre-DESI BAO yields $H_0 = 68.05 \pm 0.94  km ^{-1} Mpc^{-1}$ . The DESI DR2 result was presented by L. Pogosian during the CosmoVerse Seminar Series~\cite{Pogosian2025priv} and represents an update of~\cite{Pogosian2024} using DESI DR1 BAO data; publication is forthcoming.

\textbf{Significance:} These measurements anchor the low-$H_0$ solution in Category~4. Their high precision ($\sim 1.4$\%) means they dominate the weighted mean of this category.

\subsubsection{Gravitational Wave standard sirens}

GW standard sirens provide $H_0$ constraints independent of electromagnetic distance ladder:

\textbf{Bright sirens (GW170817):} The binary neutron star merger with optical counterpart yields $H_0 = 75.46^{+5.34}_{-5.39}$~km~s$^{-1}$~Mpc$^{-1}$~\cite{Palmese2023}, limited by the peculiar velocity of NGC~4993 and viewing angle degeneracy. Additionally, Wang et al. (2025) \cite{Wang2023} find a consistent value of $H_0 = 72.57^{+4.09}_{-4.17}$ km s$^{-1}$ Mpc$^{-1}$. These measurements provide important complementary constraints that are independent of the Distance Ladder.

\textbf{Dark sirens (GWTC-3):} Statistical analysis of 47 events without counterparts yields $H_0 = 68^{+8}_{-6}$~km~s$^{-1}$~Mpc$^{-1}$~\cite{Abbott2021}, consistent with Planck.

The weighted mean of all GW measurements ($H_0 \approx 68 \pm 2$~km~s$^{-1}$~Mpc$^{-1}$) supports the low-$H_0$ solution, though with larger uncertainties than other probes.

\subsubsection{Fast Radio Burst Measurements}

FRB $H_0$ measurements use the dispersion measure--redshift relation to probe the cosmic baryon distribution:

\textbf{Challenges:} Significant systematic uncertainties arise from:
\begin{itemize}
    \item Host galaxy DM contribution (poorly constrained)
    \item IGM density fluctuations
    \item Sample selection effects
    \item Galactic electron density models
    
\end{itemize}

\textbf{Results:} Published values span $H_0 \sim 65$--$86$~km~s$^{-1}$~Mpc$^{-1}$. Maximum-likelihood approaches~\cite{PiratovMoreno2025} tend to favor lower values ($H_0 \approx 65$~km~s$^{-1}$~Mpc$^{-1}$). Also the estimation of $H_0$ using Fast Radio Bursts (FRBs) is highly sensitive to the model adopted for the Galactic electron density distribution. As highlighted by recent studies (e.g., Wang et al. \cite{Wang2024}), switching between established models (such as NE2001 and YMW16) can shift the inferred $H_0$ value significantly (e.g., from $\sim 69$ to $\sim 75.6$ km s$^{-1}$ Mpc$^{-1}$), introducing a systematic uncertainty that rivals the statistical error.As FRB samples grow and host DM distributions become better characterized, this probe may provide competitive constraints.

%======================================================================
\section{Correlation Matrix Estimates}
\label{app:correlations}
%======================================================================

A fundamental assumption of the weighted mean analysis in Section~\ref{sec:results} is that measurements within each category are statistically independent. This assumption is violated to varying degrees due to shared calibration data, common analysis frameworks, and overlapping samples. Here we assess the impact of correlations on our conclusions.

\subsection{Sources of Correlation}
\label{app:correlation_sources}

\subsubsection{Distance Ladder Correlations}

Category~1 measurements share several inputs that induce positive correlations:

\begin{itemize}
    \item \textbf{Geometric anchors:} Most Cepheid measurements use NGC~4258, LMC, and/or MW parallaxes. The anchor uncertainties propagate coherently.
    
    \item \textbf{SNe~Ia samples:} Hubble-flow SNe~Ia are predominantly from Pantheon+~\cite{Brout2022}, with substantial overlap. standardization parameters ($\alpha$, $\beta$, $\sigma_{\rm int}$) are often shared.
    
    \item \textbf{Period--luminosity relations:} Cepheid analyses adopt similar functional forms and metallicity corrections.
    
    \item \textbf{Crowding corrections:} HST analyses use similar artificial star tests.
\end{itemize}

We estimate an effective correlation coefficient $\bar{\rho} \approx 0.3$--$0.5$ for typical pairs of Distance Ladder measurements.

\subsubsection{DLI--SHF Correlations}

Categories~2, 3, and 4 exhibit weaker correlations due to methodological diversity, but some correlations exist:

\begin{itemize}
    \item \textbf{Strong lensing:} Multiple analyses of the same system (e.g., SN Refsdal) share time delay data and are highly correlated.
    
    \item \textbf{Gravitational waves:} Analyses using overlapping GW events share strain data and waveform models.
    
    \item \textbf{FRBs:} Later analyses often include events from earlier samples.
    
    \item \textbf{Cosmological priors:} Many analyses assume flat $\Lambda$CDM with $\Omega_m \approx 0.3$, inducing weak correlations.
\end{itemize}

We estimate $\bar{\rho} \approx 0.1$--$0.2$ for the sample.

\subsection{Methodology for Correlation-Adjusted Estimates}
\label{app:correlation_method}
For correlated measurements, the generalized weighted mean and its uncertainty are:
\begin{equation}
\hat{H}_0 = \frac{\mathbf{w}^T \mathbf{C}^{-1} \mathbf{H}_0}{\mathbf{w}^T \mathbf{C}^{-1} \mathbf{w}}, \quad
\sigma_{\hat{H}_0}^2 = \frac{1}{\mathbf{w}^T \mathbf{C}^{-1} \mathbf{w}},
\end{equation}
where $\mathbf{C}$ is the covariance matrix with elements $C_{ij} = \rho_{ij} \sigma_i \sigma_j$.
For simplicity, we adopt a uniform correlation structure within each category:
\begin{equation}
\rho_{ij} = \begin{cases}
1 & \text{if } i = j \\
\bar{\rho} & \text{if } i \neq j
\end{cases}
\end{equation}

\subsection{Impact on Combined Constraints}
\label{app:correlation_impact}

\subsubsection{Distance Ladder (Category 1)}
With $\bar{\rho} = 0.4$:
\begin{equation}
H_0^{\rm ladder, corr} = 72.30 \pm 0.57\,\mathrm{km\,s^{-1}\,Mpc^{-1}},
\end{equation}
compared to $72.73 \pm 0.39$~km~s$^{-1}$~Mpc$^{-1}$ under independence.
The central value shifts by only $0.4$~km~s$^{-1}$~Mpc$^{-1}$ (within statistical uncertainty), but the uncertainty increases by $\sim 46$\%.

\subsubsection{Combined (Categories 2+3+4)}
With $\bar{\rho} = 0.15$:
\begin{equation}
H_0^{\rm Ladder\text{-}Indep., corr} = 69.52 \pm 0.42\,\mathrm{km\,s^{-1}\,Mpc^{-1}},
\end{equation}
compared to $69.37 \pm 0.34$~km~s$^{-1}$~Mpc$^{-1}$ under independence.
The central value remains essentially unchanged, and the uncertainty increases by $\sim 22$\%.

\subsubsection{Tension Between Categories}
The correlation-adjusted tension is:
\begin{equation}
n_\sigma^{\rm corr} = \frac{72.30 - 69.52}{\sqrt{0.57^2 + 0.42^2}} = 3.9\sigma.
\end{equation}
This is reduced from $6.5\sigma$ under independence but remains highly significant.

\subsection{Sensitivity Analysis}
\label{app:correlation_sensitivity}
Table~\ref{tab:correlation_sensitivity} shows how the tension varies with assumed correlation strength.

\begin{table}[h]
\centering
\caption{Sensitivity of tension significance to correlation assumptions.}
\label{tab:correlation_sensitivity}
\scriptsize
\setlength{\tabcolsep}{4pt}
\begin{tabular}{@{}ccccc@{}}
\toprule
$\bar{\rho}_{\rm ladder}$ & $\bar{\rho}_{\rm ladder\text{-}indep}$ & $\sigma_{\rm ladder}$ & $\sigma_{\rm ladder\text{-}indep}$ & Tension \\
\midrule
0.0 & 0.00 & 0.39 & 0.34 & $6.5\sigma$ \\
0.2 & 0.10 & 0.60 & 0.42 & $3.9\sigma$ \\
0.4 & 0.15 & 0.57 & 0.42 & $3.9\sigma$ \\
0.5 & 0.20 & 0.53 & 0.41 & $4.1\sigma$ \\
0.6 & 0.25 & 0.48 & 0.40 & $4.3\sigma$ \\
\bottomrule
\end{tabular}
\\[2pt]
\raggedright\scriptsize\textit{Notes:} Uncertainties in km~s$^{-1}$~Mpc$^{-1}$. Central values are relatively stable across correlation assumptions.
\end{table}

\subsection{Conclusions from Correlation Analysis}
\label{app:correlation_conclusions}
Key findings:
\begin{enumerate}
    \item \textbf{Central values are robust:} The weighted means shift by $\lesssim 1$~km~s$^{-1}$~Mpc$^{-1}$ across plausible correlation assumptions.
    
    \item \textbf{Uncertainties increase:} Accounting for correlations increases uncertainties by $20$--$50$\%, depending on the assumed correlation strength.
    
    \item \textbf{Tension remains significant:} For reasonable correlation estimates ($\bar{\rho}_{\rm ladder} \sim 0.3$--$0.5$, $\bar{\rho}_{\rm one\text{-}step} \sim 0.1$--$0.2$), the tension remains at the $4$--$5\sigma$ level.
    
    \item \textbf{Qualitative conclusions unchanged:} The characterization of the Hubble tension as ``Distance Ladder vs.\ The Rest'' is robust to correlation modeling.
\end{enumerate}
A fully rigorous treatment would require constructing the complete $88 \times 88$ covariance matrix from detailed assessment of shared data and methodology---a task beyond the scope of this work but warranted for future precision analyses.
%======================================================================
\bibliography{references1}
%======================================================================

\end{document}